\documentclass[preprint,aps,prd,numbers,sort&compress,nofootinbib,mcite]{revtex4-1}
\usepackage{graphicx}
\usepackage{float}
\usepackage{bm}
\usepackage{amssymb}
\usepackage{slashed}
\usepackage{amsmath}
\usepackage{verbatim}
\usepackage{color}
\usepackage{xcolor}
\usepackage{subfigure}
\usepackage{epstopdf}
\usepackage{multirow}
\usepackage{tabularx}
\usepackage{appendix}
\allowdisplaybreaks[4]

\usepackage{hyperref}
\hypersetup{colorlinks=true,citecolor=blue,urlcolor=blue,linkcolor=blue}

\newcounter{JW}

\begin{document}
\title{Charged-current non-standard neutrino interactions at the LHC and HL-LHC}

\author{Chong-Xing Yue\textsuperscript{1,2}}
\email{cxyue@lnnu.edu.cn}
\author{Xue-Jia Cheng\textsuperscript{1,2}}
\email{cxj225588@163.com}
\author{Ji-Chong Yang\textsuperscript{1,2}}
\email{yangjichong@lnnu.edu.cn}
\date{\today}
\affiliation{\textsuperscript{1}Department of Physics, Liaoning Normal University, Dalian 116029, China\\\textsuperscript{2}Center for Theoretical and Experimental High Energy Physics, Liaoning Normal University, Dalian 116029, China}

\begin{abstract}
A series of new physics scenarios predict the existence of the extra charged gauge boson $W'$, which can induce charged-current~(CC) non-standard neutrino interactions~(NSIs).
The theoretical constraints on the simplified $W'$ model and further on the CC NSI parameters $\widetilde{\epsilon}^{ qq'Y}_{\alpha\beta}$ from partial wave unitarity and $W'$ decays are considered. The sensitivity of the process $p p \rightarrow W'\rightarrow \ell\nu $ to the $W'$ model at the LHC and high-luminosity~(HL) LHC experiments is investigated by estimating the expected constraints on $\widetilde{\epsilon}^{qq'Y}_{\alpha\beta}$ ($\alpha = \beta = e $ or $\mu$) using a Monte-Carlo~(MC) simulation.
We find that the interference effect plays an important role, and the LHC can strongly constrain $\widetilde{\epsilon}^{qq'L}_{\alpha\beta}$.
Compared with those at the $13 \;{\rm TeV}$ LHC with  $\mathcal{L}=139\;{\rm fb}^{-1}$, the expected constraints at the $14 \;{\rm TeV}$ LHC with  $\mathcal{L}=3\;{\rm ab}^{-1}$ can be strengthened to approximately one order of magnitude.
\end{abstract}

\maketitle

\section{\label{level1} Introduction }
The Standard Model~(SM) has achieved great success in describing elementary particles and interactions.
Numerous experimental studies have verified its predictions with very high accuracy.
However, the SM does not reveal the origin of the neutrino mass.
Strong evidence show that neutrinos with different flavors cannot oscillate with each other without small mass differences.
The origin of neutrino mass requires new physics beyond the SM~(BSM)~\cite{Barger:2003qi,Gonzalez-Garcia:2007dlo}, many examples of which share the common feature of emerging effective non-standard neutrino interactions~(NSIs) between neutrinos and matter fields~\cite{Wolfenstein:1977ue,Choudhury:2018xsm,Ohlsson:2012kf,Miranda:2015dra,Farzan:2017xzy}.
NSIs are generally divided into two types: the neutral-current~(NC) NSI and charged-current~(CC) NSI.
The NC NSI mainly affects neutrino propagation in matter, whereas the CC NSI is associated with neutrino production and detection processes (for recent reviews, see Refs.~\cite{Ohlsson:2012kf,Miranda:2015dra,Farzan:2017xzy,Esteban:2018ppq}).
These new interactions lead to rich phenomenology in neutrino oscillation experiments and high- and low-energy collider experiments.
Very strong constraints on the NSIs have been obtained, see, for example, Refs.~\cite{Choudhury:2018xsm,Gonzalez-Garcia:2011vlg,Gonzalez-Garcia:2013usa,Davidson:2011kr,Friedland:2011za,BuarqueFranzosi:2015qil,Babu:2019mfe,Terol-Calvo:2019vck,Escrihuela:2011cf,Altmannshofer:2018xyo,Yue:2022eac}.
The experiments generally impose more strict restrictions on the CC NSI than on the NC NSI ~\cite{Davidson:2003ha,Biggio:2009nt}.

Recently, numerous phenomenological studies have been conducted on NSIs.
For example, the new gauge boson $Z'$ can introduce NC NSIs~\cite{Cheung:2021tmx,Pandey:2019apj,Friedland:2011za,Liu:2020emq,Babu:2020nna,Bischer:2018zbd,Coloma:2020gfv}.
Refs.~\cite{Friedland:2011za,Pandey:2019apj,Liu:2020emq,Babu:2020nna} studied the constraints on NC NSIs in the context of a simplified $Z'$ model using mono-jet signals at the Large Hadron Collider~(LHC).
Many BSMs predict the existence of $W'$, such as the left-right symmetry model~\cite{Mohapatra:1974hk,Senjanovic:1975rk}, little Higgs models~\cite{Arkani-Hamed:2002ikv}, and models with extra dimensions~\cite{Appelquist:2000nn,Agashe:2008jb}, which can lead to CC NSIs.
Previous searches for the $W'$ boson at the LHC have been carried out by the ATLAS and CMS collaborations with data collected at the center-of-mass (c. m.) energy $\sqrt{s}$ = $7 \;{\rm TeV}$~\cite{ATLAS:2012aqf,CMS:2012mwz}, $8\;{\rm TeV}$~\cite{ATLAS:2014wra,CMS:2014fjm}, and $13 \;{\rm TeV}$~\cite{ATLAS:2017jbq,CMS:2016ifc}.
The most sensitive channels are $e\nu$ and $\mu\nu$ production, with the constraints to date set in Refs.~\cite{CMS:2018hff,ATLAS:2019lsy}, which can be translated to constraints on CC NSIs.
The LHC experiments have further promoted the theoretical research on $W'$ properties~\cite{Osland:2020onj,Fiaschi:2021sin,Calabrese:2021lcz}.
Although there are other new particles that can also cause the CC NSI, such as charged scalar particles (for example, see~\cite{Babu:2019mfe}), in this paper, we focus on the simplified $W'$ model.

It has been shown that there is a well-known `degeneracy' between the parameter spaces of NSIs in neutrino oscillation experiments because the effects of NSIs strongly depend on the flavor structure and oscillation channel being studied~\cite{Coloma:2019mbs}.
Although it is difficult to break the degeneracy at neutrino facilities, the LHC plays a complementary role in the study of NSIs~\cite{Choudhury:2018xsm,Davidson:2011kr,Proceedings:2019qno,Babu:2019mfe}, and it is believed to be able to break the degeneracy because the effects of NSIs at the LHC do not distinguish between different neutrino flavors~\cite{Liu:2020emq,Babu:2020nna}.
Furthermore, the LHC is sensitive to both vector- and axial-vector interactions, whereas neutrino oscillation experiments are not sensitive to the latter.

Thus far, various studies have been completed on the detection of $W'$ at the LHC, and searching for this type of new particle will continue to be important at the future LHC with higher luminosities (HL-LHC).
Based on a Monte-Carlo~(MC) simulation, the expected constraints on the CC NSI parameters via the process $pp\to \ell^+  \nu_{\ell} $ are studied with an emphasis on the effects of the interference between the CC NSI and the SM. Moreover, the sensitivities at future runs of the LHC are also discussed.

The remainder of this paper is organized as follows:
In Sec.~\ref{level2}, we consider the theoretical constraints on the simplified $W'$ model and further on the CC NSI parameters $\widetilde{\epsilon}^{qq'Y}_{\alpha\beta}$ ($\alpha=\beta=e$ or $\mu$) in terms of partial wave unitarity and $W'$ decay.
The sensitivities of the process $pp\to \ell^+ \nu_{\ell}$ to the parameters $\widetilde{\epsilon}^{udY}_{\alpha\beta}$ at the current LHC and future runs of the LHC are studied in Secs.~\ref{level3} and~\ref{level4}, respectively. Our conclusions and discussions are given in Sec.~\ref{level5}.

\section{\label{level2}Theoretical constraints on the simplified \texorpdfstring{$W'$}{W'} model and NSI parameters }

NSIs are new vector interactions between neutrinos and matter fields, induced by either a vector or scalar mediator. They can be parameterized in terms of the low-energy effective four-fermion Lagrangian~\cite{Davidson:2003ha,Biggio:2009nt,Proceedings:2019qno,Biggio:2009kv}.
CC NSIs with quarks are given by the effective Lagrangian
\begin{equation}
\mathcal{L}_{NSI,CC} =  -2{\sqrt{2}}{G_{F}}\epsilon^{qq'Y}_{\alpha\beta}\left[\bar q \gamma^{\mu}P_{Y}q'\right]\left[\bar{\nu_{\alpha}}\gamma_{\mu}P_{L}\ell_{\beta}\right]+h.c.,
\label{eq.antsaz1}
\end{equation}
where ${G_{F}}$ is the Fermi constant, $\alpha$ and $\beta$ are lepton flavor indices, $\alpha,\beta \in \{e,\mu,\tau\}$, and $P_{Y}$ is a chiral projection operator ($P_{L}$ or $P_{R}$).
We assume that there are only left-handed neutrinos in the above equation.
The parameters $\epsilon^{qq'Y}_{\alpha\beta}$ are dimensionless coefficients that quantify the strengths of the new vector interactions.
According to Hermiticity, $\epsilon^{qq'Y}_{\alpha\beta}$ = $\epsilon^{qq'Y\ast}_{\beta\alpha}$.
The CC NSI might change the production of neutrinos through its effects on processes such as muon decay and inverse beta decay~\cite{Soumya:2019kto,Santos:2020dgs}.

Because the momentum transfer can be very high to resolve further dynamics of new physics at the LHC, the influence of the CC NSI may not be simply described by the effective operators.
In this paper, we focus on a simplified model with the CC NSI induced by the exchange of a $W'$ boson.
The effective Lagrangian can be written as~\cite{Sullivan:2002jt,Gopalakrishna:2010xm,Fuks:2017vtl}
\begin{equation}
\mathcal{L}_{W'} =
-\frac{g}{\sqrt{2}}\left[V_{qq'}\bar{q}\gamma^{\mu}\left({A}^{qq'}_{L}P_{L}+{A}^{qq'}_{R}P_{R}\right)q'+{B}^{\alpha\beta}_{L}\bar{\ell_{\alpha}}\gamma^{\mu}P_{L}\nu_{\beta}\right]{W}^{'}_{\mu}+h.c.,
\label{eq.antsaz2}
\end{equation}
where ${W}^{'}_{\mu}$ denotes the new force carrier with a mass ${M}_{W'}$, and $g$ is the electroweak coupling constant.
$q= \{u,c,t\}$ and $q'= \{d,s, b\}$ are up-type and down-type quarks, respectively.
$V_{qq'}$ is the Cabbibo-Kobayashi-Maskawa(CKM) matrix element, in which the non-diagonal terms contribute to flavor changing neutral current (FCNC) processes~\cite{Branco:2013evr}, and the contributions are very small and negligible.
For simplicity, one can also absorb $g V_{qq'}/ \sqrt{2}$ into the coupling parameters $A^{qq'}_{Y}$ and $B^{\alpha\beta}_{L}$.

In a simplified model framework, the relationship between Eq.~(\ref{eq.antsaz1}) and Eq.~(\ref{eq.antsaz2}) can be thought of as $\epsilon^{qq'Y}_{\alpha\beta} \sim g^{2}_{X} / M^{2}_{X}$, with $X$ representing a charged mediator. Because the Wilson coefficients of an effective field theory (EFT) are typically functions of energy scales, Eq.~(\ref{eq.antsaz2}) implies a substitution of $\epsilon^{qq'Y}_{\alpha\beta}$ in Eq.~(\ref{eq.antsaz1}), $\epsilon^{qq'Y}_{\alpha\beta} \to \epsilon^{qq'Y}_{\alpha\beta} (s)$, with $\epsilon^{qq'Y}_{\alpha\beta} (s) = -M_{W'}^2\widetilde{\epsilon}^{qq'Y}_{\alpha\beta} / \left(s-M_{W'}^2\right)$, and $\epsilon^{qq'Y}_{\alpha\beta} = \widetilde{\epsilon}^{qq'Y}_{\alpha\beta}$ when $s \to 0$, and
\begin{equation}
\widetilde{\epsilon}^{qq'L}_{\alpha\beta} =
{A}^{qq'}_{L} {B}^{\beta\alpha\ast}_{L}\left(\frac{M_{W}}{M_{W'}}\right)^{2},\;\;
\widetilde{\epsilon}^{qq'R}_{\alpha\beta} =
{A}^{qq'}_{R}{B}^{\beta\alpha\ast}_{L}\left(\frac{M_{W}}{M_{W'}}\right)^{2}.
\label{eq.antsaz3}
\end{equation}
However, it can also be considered that the effective operator in Eq.~(\ref{eq.antsaz1}) corresponds to the leading order of expansion of $\epsilon^{qq'Y}_{\alpha\beta} (s)$ in $s/M_{W'}^2$, whereas Eq.~(\ref{eq.antsaz2}) also collects the contributions of the higher dimensional operators corresponding to the higher orders of $s/M_{W'}^2$.

In principle, there are also flavor violating NSI couplings. Neutrino oscillation and scattering experiments are found to have tight constraints on diagonal NSIs, whereas off-diagonal NSIs are primarily constrained by various low-energy processes, such as atomic parity violation and charged-lepton flavor violation~(cLFV)~\cite{Babu:2019mfe}. Because only diagonal NSIs are relevant for the process $pp\to \ell \nu$, we concentrate on the $\alpha=\beta$ case in this paper.

\subsection{ Unitarity constraints from the \texorpdfstring{$f\bar{f}\to V_1V_2$}{ff to V1 V2} process}

\begin{figure}[H]
	\centering
			\includegraphics [width=4.7cm] {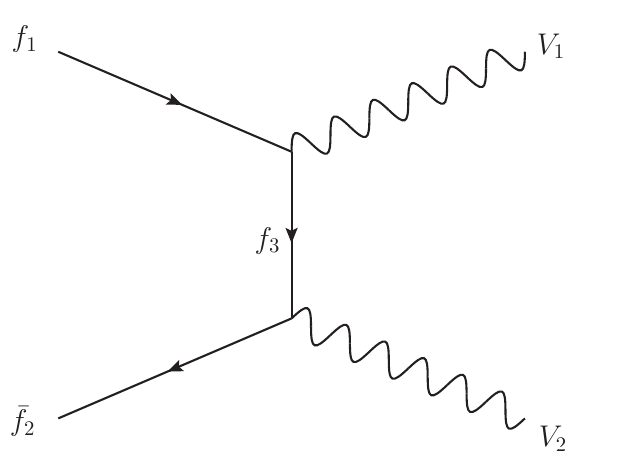}
			\includegraphics [width=5cm] {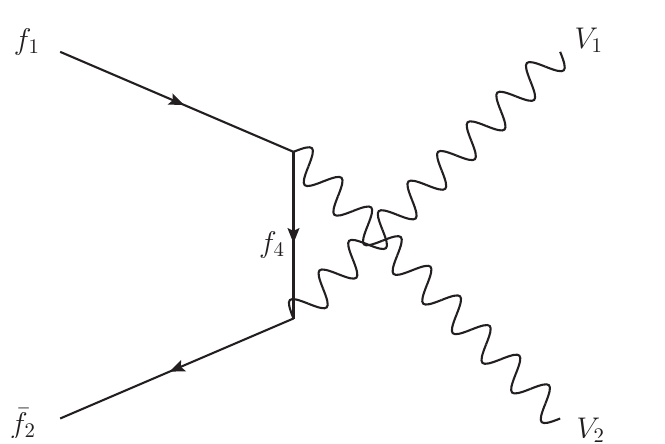}
			\includegraphics [width=4.8cm] {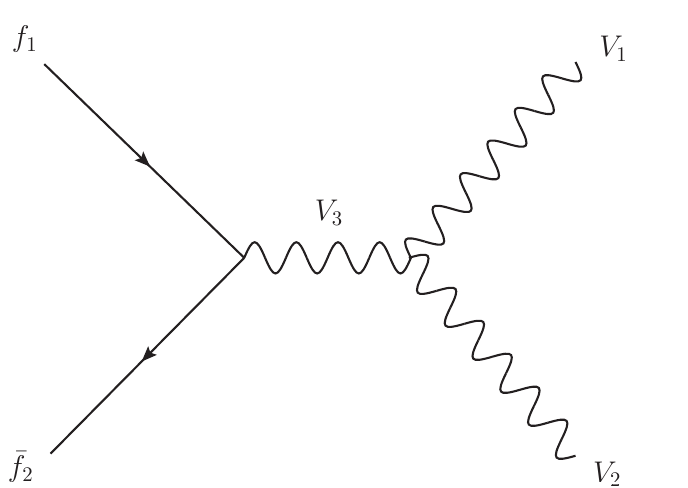}
	\caption{\label{fig:20} Tree-level Feynman diagrams of the process $f\bar{f}\to V_1V_2$.   }
\end{figure}

In a weakly coupled renormalizable theory, the high-energy behavior of the scattering amplitude of bosons under the perturbation calculation is expected to respect unitarity.
It has been shown that the constraint on $M_{Z'}$ can be obtained using partial wave unitarity bounds derived from the $f\bar{f}\to V_1V_2$ process, where $V$ represents vector bosons, including SM $W,Z$ bosons and also the new gauge bosons $W',Z'$, and $f$ and $\bar{f}$ are fermions and anti-fermions~\cite{Babu:2011sd}.
The helicity amplitudes can be expanded as~\cite{Jacob:1959at,Baur:1987mt}
\begin{equation}
\mathcal{M}_{f_{\sigma _1}\bar{f}_{\sigma _2}\to V_{1,\lambda _3}V _{2,\lambda _4}}
 =16\pi\sum _J \left(J+\frac{1}{2}\right)\delta _{\sigma_1,-\sigma _2}e^{i(m_1-m_2)\phi}d^J_{m_1,m_2}(\theta,\phi) T_J,
\label{eq.amplitude}
\end{equation}
where $m_1=\sigma _1-\sigma _2$ and $m_2=\lambda _3-\lambda _4$ are the helicity differences of fermions and vector bosons, respectively. $d^J_{m_1,m_2}$ are Wigner $d$-functions~\cite{Jacob:1959at}, $\phi$ and $\theta$ are the azimuth and zenith angles of $V_1$ in the rest frame of the fermions whose $Z$-axis points in the direction of $f$. The partial wave unitarity bound is $|T_J|\leq 1$~\cite{Corbett:2014ora,Fu:2021mub}, where $T_J$ is the coefficient of partial wave expansion. As noted in Ref.~\cite{Babu:2011sd}, we only need to consider processes with the longitudinal final states $f_{\pm \frac{1}{2}}\bar{f}_{\pm \frac{1}{2}}\to V_{1,\lambda_3=0}V_{2,\lambda_4=0}$, whose amplitudes are denoted as $\mathcal{M}^{( - +)}_{f_{1}\bar{f}_{2}\to V_{1}V _{2}}$ and $\mathcal{M}^{( + -)}_{f_{1}\bar{f}_{2}\to V_{1}V _{2}}$, where `$-+$' in superscript means $f_{- \frac{1}{2}}\bar{f}_{ +\frac{1}{2}}$, and `$+-$' means $f_{+\frac{1}{2}}\bar{f}_{-\frac{1}{2}}$.
In the following, we only consider tree-level processes, including t-channel, u-channel, and s-channel processes. The corresponding Feynman diagrams are shown in Fig.~\ref{fig:20}. There is also an s-channel scalar exchange diagram, which only contributes to $\mathcal{M}^{(+ +)}_{f_{1}\bar{f}_{2}\to V_{1}V _{2}}$ and $\mathcal{M}^{(- -)}_{f_{1}\bar{f}_{2}\to V_{1}V _{2}}$; therefore, it is not shown.

$\mathcal{M}^{( - +)}_{f_{1}\bar{f}_{2}\to V_{1}V _{2}}$ and $\mathcal{M}^{( + -)}_{f_{1}\bar{f}_{2}\to V_{1}V _{2}}$ can be expanded in terms of $s$ as
\begin{equation}
\mathcal{M}^{(\pm\mp)}_{f_{1}\bar{f}_{2}\to V_{1}V _{2}}
 = \mathcal{A}_{1} s + \mathcal{A}_{0} + \mathcal{A}_{-1} s^{-1} + \mathcal{A}_{-2} s^{-2} + \cdots.
\end{equation}
Partial wave unitarity should be satisfied for arbitrary $s$. When $s\to \infty$, a finite $|T_J|$ implies $\mathcal{A}_{1} = 0$, which results in a set of `sum rules'~\cite{Babu:2011sd,Romao:2012pq},
\begin{equation}
\sum_{f_{3}} g^{L/R}_{V_1 f_1 \bar f_3}g^{L/R}_{V_2 f_3 \bar f_2} - \sum_{f_{4}} g^{L/R}_{V_1 f_4 \bar f_2}g^{L/R}_{V_2 f_1 \bar f_4} = \sum_{V_{3}} g^{L/R}_{V_3 f_1 \bar f_2}g^{L/R}_{V_1 V_2 V_3},
\label{eq.sumrule}
\end{equation}
where $V_{1,2,3}$ and $f_{1,2,3,4}$ are allowed vector bosons and fermions in the Feynman diagrams in Fig.~\ref{fig:20}, respectively.

For the scattering amplitudes of the process $f_{1}\bar{f}_{2}\to V_{1}V _{2}$, a new gauge boson $Z'$ must have non-zero coupling with fermions and other gauge bosons, and the existence of $Z'$ is necessary to maintain the perturbative unitarity of all amplitudes.
Therefore, triple gauge couplings involving $Z'$ and $Z'$-fermion couplings must be involved in the calculation of the Feynman diagrams shown in Fig.~\ref{fig:20}.  There are enough `sum rules' to solve all unknown couplings of $Z'$ using Eq.~(\ref{eq.sumrule}), that is, the $Z'$-fermion couplings are automatically fixed when the $W'$-fermion couplings are fixed~\cite{Babu:2011sd}.
Certainly, the $W-W'$ and $Z-Z'$ mixings would modify the relevant SM couplings slightly, which can affect the amplitudes of the process $f_{1}\bar{f}_{2}\to V_{1}V _{2}$. It has been shown that the correction effect is less than $1\%$~\cite{Babu:2011sd,Electroweak:2003ram}; hence, for simplicity, we neglect it in our calculations.
With the help of Eq.~(\ref{eq.sumrule}), the amplitudes corresponding to the Feynman diagrams in Fig.~\ref{fig:20} are written as Eq.~(\ref{eq.amplitudes}).

Using the relationship $|T_J|\leq 1$, one can obtain a constraint on ${A}^{qq'}_{Y}$ or ${B}^{\alpha\alpha}_{L}$ from each of the above processes. All of these constraints should be satisfied, therefore, we concentrate on the tightest ones, which depend on the free parameters $M_{W'}$ and $M_{Z'}$. In the case of $0.2\;{\rm TeV}<M_{Z'}<2\;{\rm TeV}$~\cite{Liu:2020emq} and $s\to \infty$, the tightest bounds on ${A}^{qq'}_{Y}$ or ${B}^{\alpha\alpha}_{L}$ are given in Eq.~(\ref{eq.coupling}). The relationship between $M_{W'}$ and the coupling parameters according to Eq.~(\ref{eq.coupling}) are shown in Fig.~\ref{figa}. It can be seen that in the wide range of $M_{W'}$, the tightest bound on ${A}^{qq'}_{L}$ originates from the process $u\bar{d}~(d\bar u)\to W'Z$. Similarly, the tightest bound on ${B}^{\alpha\alpha}_{L}$ mainly arises from the process $e^-e^+\to W'Z$. For simplicity, we only use the above bounds.
For ${A}^{qq'}_{R}$, the two tightest bounds~(depending on $M_{W'}$) are both considered.
Consequently, the perturbative unitarity constraints on the parameters $\widetilde{\epsilon}^{qq'Y}_{\alpha \alpha}$ are
\begin{equation}
\begin{aligned}
&\left|\widetilde{\epsilon}^{qq'L }_{\alpha\alpha}\right| < \frac{1152\pi^{2} M^{2}_{Z}M^{2}_{W} }{g^{4}c^{2}_{W}  \left|V_{qq'}\right|^{2} \left(2 M_{W'}^{2}+M_{Z}^{2}\right)^{2}},\\
&\left|\widetilde{\epsilon}^{qq'R }_{\alpha\alpha}\right| < {\rm min} \left\{ \frac{1152\pi^{2} M^{2}_{Z}M^{2}_{W}}{ \left|V_{qq'}\right|^{2} s^{2}_{W} g^{4}\left(2 M_{W'}^{2}+M_{Z}^{2}\right)^{2}}\right.,\\
&\left.\frac{48\sqrt{2}\pi M_{Z} M^{2}_{W}}{g^{2}c_{W}M_{W'}\left(2 M_{W'}^{2}+M_{Z}^{2}\right)}\sqrt{\frac{72\sqrt{2}\pi c^{2}_{W} M^{2}_{W'}+ s^{2}_{W}g^{2}\left|V_{qq'}\right|^{2}\left(s^{2}_{W} M^{2}_{Z}-M^{2}_{Z'}\right)}{3c^{2}_{W} g^{2}\left|V_{qq'}\right|^{2}\left(4M^{2}_{W'}-M^{2}_{Z'}\right)}} \right\}.
\label{eq.boundY}
\end{aligned}
\end{equation}

By using Eq.~(\ref{eq.antsaz1}), the unitarity bounds on $\epsilon ^{qq'Y}_{\alpha \alpha}$ can be directly obtained from the process $f\bar{f}\to f\bar{f}$. However, because the $W'$ model corresponds to  the substitution $\epsilon ^{qq'Y}_{\alpha \alpha}\to \epsilon ^{qq'Y}_{\alpha \alpha}(s)$, where the latter is just the `form factor unitarization' widely used in the study of SM EFT ~\cite{ff1,ff2,ff3}, that is, unitarity is guaranteed with this substitution; therefore, the unitarity bounds from the process $f\bar{f}\to f\bar{f}$ are not considered.

\subsection{Constraints from \texorpdfstring{${W'}$}{W'} decays}

\begin{figure}
	\begin{center}
		\includegraphics[width=0.6\textwidth]{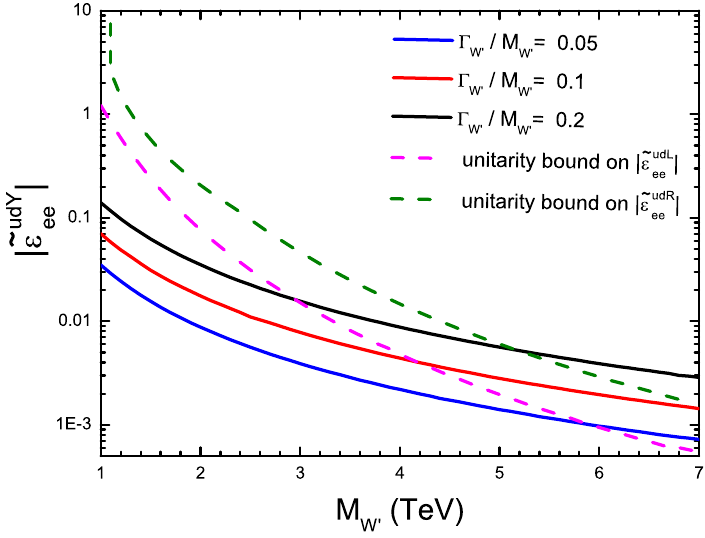}
		\caption{Constraints on the CC NSI parameter $\left|\widetilde{\epsilon}^{udY}_{ee}\right|$ from partial wave  unitarity and $W'$ decays for different values of the ratio $\Gamma _{W'}/M_{W'}$.}
		\label{fig:1}
	\end{center}
\end{figure}

From Eq.~(\ref{eq.antsaz2}), the total decay width of the $W'$ boson at leading order can be written as
\begin{equation}
\Gamma_{W'} =
M_{W'}\left(\frac{{g}^{2}}{16\pi}\sum_{q,q'}\left|V_{qq'}\right|^2\left(|A^{qq'}_{L}|^{2}+|A^{qq'}_{R}|^{2}\right)+\frac{{g}^{2}}{48\pi}\left(\left|B_{L}^{ee}\right|^{2}+
\left|B_{L}^{\mu\mu}\right|^{2}+\left|B_{L}^{\tau\tau}\right|^{2}\right)\right).\\
\end{equation}
Without loss of generality, using the case of $A^{ud}_{L}\neq 0, B_{L}^{ee}\neq 0$ as an example, when the mass of fermions are negligible compared with $M_{W'}$,
\begin{equation}
\begin{split}
&\frac{\Gamma_{W'}}{M_{W'}} =\sum _{qq',Y}\frac{{g}^{2}|V_{qq'}|^2}{16\pi} \left|A^{qq'}_{Y}\right|^2+\frac{{g}^{2}}{48\pi}\sum _{\ell}|B_{L}^{\alpha\alpha}|^2 \geq \frac{{g}^{2}|V_{ud}|^2}{16\pi} \left|A^{ud}_{L}\right|^2 + \frac{{g}^{2}}{48\pi}|B_{L}^{ee}|^2\\
&=  \left(\sqrt{\frac{{g}^{2}|V_{ud}|^2}{16\pi}}A^{ud}_{L}-\sqrt{\frac{{g}^{2}}{48\pi}}B_{L}^{ee}\right)^{2}+2\sqrt{\frac{{g}^{4}|V_{ud}|^{2}}{768\pi^{2}}}A^{ud}_{L}B_{L}^{ee} \geq \frac{{g}^{2}V_{ud}}{8\sqrt{3}\pi }A^{ud}_{L}B_{L}^{ee}.\\
\end{split}
\end{equation}
The first inequality takes the equals sign at $A^{qq'\neq ud}_{L}=A^{qq'}_{R}=B_{L}^{\alpha\alpha\neq ee}=0$, whereas the second inequality takes the equals sign at $\sqrt{{g}^{2}|V_{ud}|^2/16\pi}A^{ud}_{L}=\sqrt{{g}^{2}/48\pi} B_{L}^{ee}$\footnote{The equal sign can be taken only in the sense of mathematics, which corresponds to the loosest case. There is no physical motivation behind it. For a physical scenario this constraint should be even tighter.}.
These inequalities hold for all $A^{qq'}_{L,R}$ and $B_{L}^{\ell\ell}$ and set an upper bound on $\widetilde{\epsilon}^{qq' Y }_{\alpha\alpha}\times M_{W'}^2$ when $\Gamma _{W'}/ M_{W'}$ is fixed,
\begin{equation}
\left|\widetilde{\epsilon}^{qq' Y }_{\alpha\alpha}\right| \leq
\frac{8\sqrt{3}\pi {M^{2}_{W}}}{{g}^{2}\left|V_{qq'}\right|{M^{2}_{W'}}}\frac{\Gamma_{W'}}{M_{W'}}.
\label{eq.theoraticalConstraint}
\end{equation}
For simplicity, taking $\left|\widetilde{\epsilon}^{udY}_{ee}\right|$ as an example, the maximally allowed $\left|\widetilde{\epsilon}^{udY}_{ee}\right|$ as a function of $M_{W'}$ is shown in Fig.~\ref{fig:1} for different values of the ratio $\Gamma _{W'}/ M_{W'}$.
The unitarity bounds given by Eq.(\ref{eq.boundY}) are also shown in Fig.~\ref{fig:1}. It can be seen that with additional assumptions on $\Gamma _{W'}/M_{W'}$, that is, for some values of $\Gamma _{W'}/M_{W'}$ and $M_{W'}$, tighter constraints can be obtained. For larger $M_{W'}$, the perturbative unitarity constraints become stronger than those from $W'$ decays. If we take $\Gamma _{W'}/ M_{W'}=0.1$~\cite{Babu:2020nna,Liu:2020emq}, $\left|\widetilde{\epsilon}^{qq' R }_{\alpha\alpha}\right|\leq 10.868\times M_W^2/M_{W'}^2$ for $1 \;{\rm TeV} \leq M_{W'} \leq 7\;{\rm TeV}$, and $\left|\widetilde{\epsilon}^{qq' L }_{\alpha\alpha}\right|\leq 10.868\times M_W^2/M_{W'}^2$ for $M_{W'} \leq 4\;{\rm TeV}$. Meanwhile, the perturbative unitarity constraints are $\left|\widetilde{\epsilon}^{qq' L }_{\alpha\alpha}\right|\leq 7.584\times M_W^2/M_{W'}^2$, $\left|\widetilde{\epsilon}^{qq' L }_{\alpha\alpha}\right|\leq 3.637\times M_W^2/M_{W'}^2$, and $\left|\widetilde{\epsilon}^{qq' L }_{\alpha\alpha}\right|\leq 2.089\times M_W^2/M_{W'}^2$ for $M_{W'} = 5\;{\rm TeV}$, $6\;{\rm TeV}$, and $7\;{\rm TeV}$, respectively.

The theoretical constraints given in Fig.~\ref{fig:1} are used as a reference for the MC studies in the next section. It can be seen from Fig.~\ref{fig:1} that the larger the mass of the $W'$ boson, the tighter the constraint on $\left|\widetilde{\epsilon}^{udL}_{ee}\right|$. However, with fixed $\Gamma _{W'}/ M_{W'}$, the $W'$ contributions to  the process $pp\to e^+\nu_{e}$ typically decrease with increasing $M_{W'}$; therefore, the larger the value of $M_{W'}$, the larger the value of $\left|\widetilde{\epsilon}^{udL}_{ee}\right|$ needed to observe the signal at the LHC. However, $\left|\widetilde{\epsilon}^{udL}_{ee}\right|$ must satisfy the unitarity constraint given by Eq.~(\ref{eq.boundY}). In the following MC analysis, we take into account the theoretical constraints on the NSI parameters $\left|\widetilde{\epsilon}^{qq' Y }_{\alpha\alpha}\right|$,  set $\Gamma _{W'}/ M_{W'}=0.1$, and assume that the mass $M_{W'}$ is from $1 \;{\rm TeV}$ to $7 \;{\rm TeV}$.

\section{\label{level3} Expected constraints on \texorpdfstring{$\widetilde{\epsilon}^{udY}_{\ell\ell}$}{epsilon} at the LHC}

\begin{figure}[H]
\begin{center}
\includegraphics[width=0.4\textwidth]{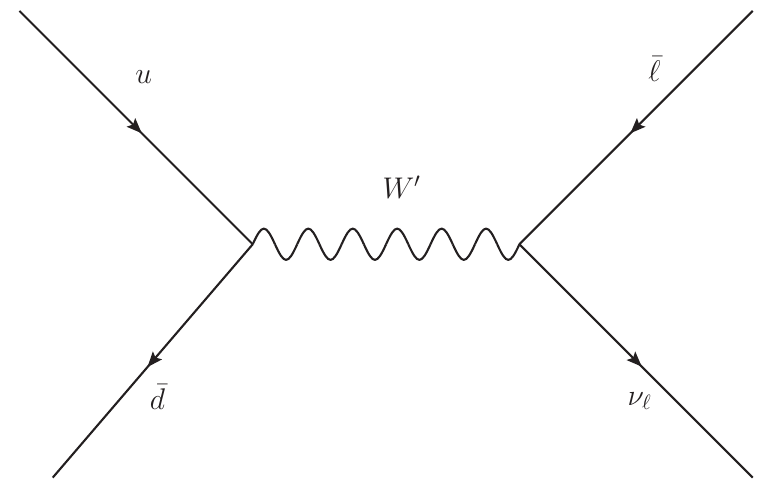}
\caption{Leading-order (LO) Feynman diagram for the single production of the $W'$ boson and subsequent leptonic decays.}
\label{fig:2}
\end{center}
\end{figure}

A simple and efficient way to search for the gauge boson $W'$ at the LHC is through its single s-channel resonance and subsequent leptonic decays~\cite{Serenkova:2019zav}.
Because the main components of protons are $u$ and $d$ quarks, the contribution of the $u \bar{d}$ process is considerably larger than that of the $c\bar{s}$ process.
It has been noted that the luminosity of $u \bar{d}$ quarks is two orders of magnitude larger than that of $c\bar{s}$ quarks~\cite{Fuentes:2020lea}. If $\widetilde{\epsilon}^{cs Y}_{\ell\ell}$ are at the same magnitude as $\widetilde{\epsilon}^{ud Y}_{\ell\ell}$, the contribution of $c\bar{s}\to \ell^{+} \nu_{\ell}$ is negligible compared with $u \bar{d} \rightarrow \ell^{+} \nu_{\ell}$; therefore, we neglect the contribution from non zero $\widetilde{\epsilon} ^{qq' Y}_{\ell\ell}$ with $qq'\neq ud$.
In this paper, we consider the subprocess $u \bar{d} \rightarrow \ell^{+} \nu_{\ell}$ where $\ell$ is $e$ or $\mu$, which corresponds to $\widetilde{\epsilon} ^{ud Y}_{ee}$ or $\widetilde{\epsilon} ^{ud Y}_{\mu\mu}$.
Fig.~\ref{fig:2} shows the Feynman diagram of the subprocess $u \bar{d} \rightarrow {\ell}^{+} \nu_{\ell}$ ($\ell= e$ or $\mu $) induced by $W'$ exchange at the LHC.

In numerical estimation, the Lagrangian described by Eq.~(\ref{eq.antsaz2}) is implemented using \verb"FEYNRULES"~\cite{Alloul:2013bka,Christensen:2008py,Degrande:2011ua}, and a Universal Feynrules Output (UFO) file is generated and taken into the \verb"MadGraph5_aMC@NLO"
toolkit~\cite{Alwall:2014hca,Sjostrand:2014zea} with the standard cuts
\begin{center}
$p_{T}^{\ell}> 10$ GeV,   $ |\eta_{\ell}| < 2.5$.
\end{center}

\begin{figure}[H]
\centering{
	\includegraphics [width=0.7\textwidth] {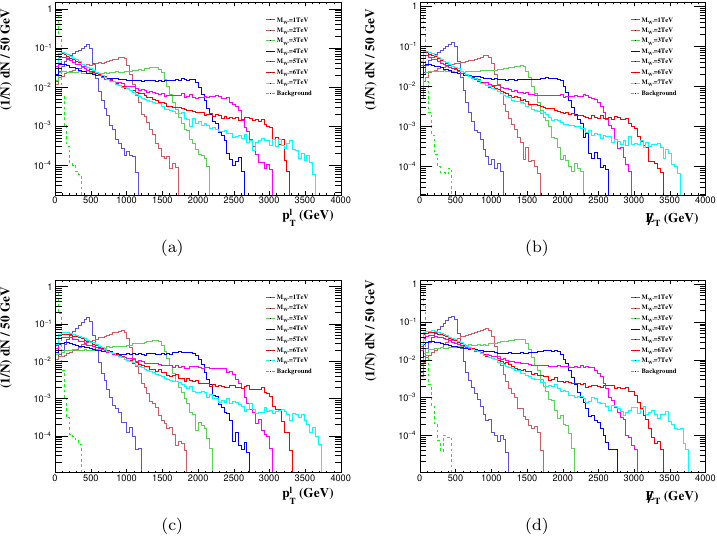}
\caption{\label{fig:4}For the process $pp \to e^{+} \nu_{e}$, the normalized distributions of $p_{T}^{\ell}$ and ${E\mkern-10.5mu/}_{T}$ for  the signal and background events. The cases for $A_{R}^{ud}$ = 0 are shown in (a,b), while the cases for $A_{L}^{ud}$ = 0 are shown in (c,d).  }}
\end{figure}

\begin{figure}[H]
\centering{
   \includegraphics [width=0.7\textwidth] {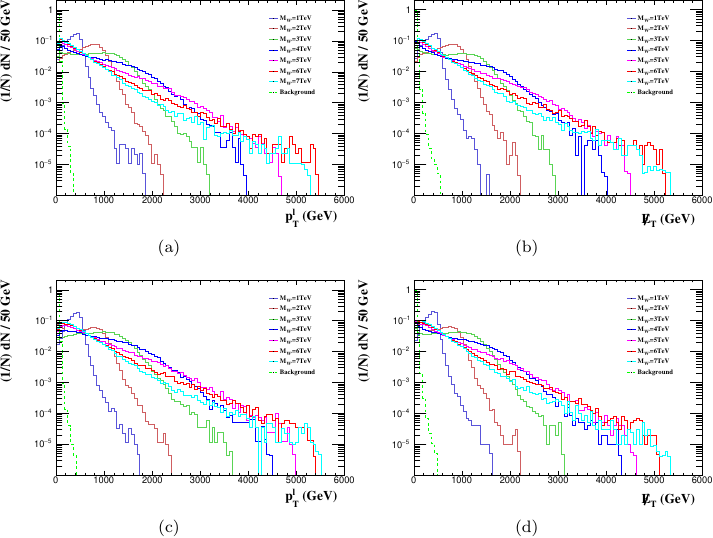}
\caption{\label{fig:5} Same as Fig.~\ref{fig:4} but for the process $pp \to \mu^{+} \nu_{\mu}$. }}
\end{figure}

A fast detector simulation is applied using \verb"Delphes"~\cite{deFavereau:2013fsa} with the CMS detector card.
The parton distribution functions are taken as  \verb"NNPDF2.3"~\cite{Ball:2011mu,Ball:2012cx}.
To highlight the signal from the background, the kinematical features of the signal and background are studied using \verb"MadAnalysis5"~\cite{Conte:2012fm}.
The processes in the SM with the same final states are considered as the background.
At tree level, there is only one type of background in which $\ell^+\nu_{\ell}$ are from the SM $W$ boson.
The main difference between the signal and the background is that for signal events, the leptons are from a $W'$ boson with a considerably lager mass $M_{W'}$ than that of the SM $W$ boson.
As a consequence, the transverse momenta of charged leptons for signal events are generally larger than those for background events.
$p^{\ell}_T$ has also been used in previous studies to highlight the signal of $W'$.
The normalized distributions of $p^{\ell}_T$ for $pp\to e^+\nu _e$ are shown in Fig.~\ref{fig:4}.~(a).
It can be seen that $p^{\ell}_T$ is generally smaller than $360\;{\rm GeV}$ for background events, whereas the signal events have large $p^{\ell}_T$ tails.
In this paper, we only keep events with $p^{\ell}_T \geq 360\;{\rm GeV}$.
For the same reason, the missing transverse energy $\slashed{E}_T$ for the signal events is also typically larger than those for the background events.
The normalized distributions of $\slashed{E}_T$ are shown in Fig.~\ref{fig:4}.~(b).
The background events are dominantly distributed in the region $\slashed{E}_T<460\;{\rm GeV}$, which is not the case for the signal events.
In this paper, we only keep events with $\slashed{E}_T \geq 460\;{\rm GeV}$.
The normalized distributions for the process $pp \to \mu^{+} \nu_{\mu}$ are shown in Fig.~\ref{fig:5}.
When searching for the signal of the $W'$ boson with an unknown mass, the cuts are applied uniformly.
However, the event selection strategies can be further improved.
If the signal is not found, the goal should be to constrain the parameter $\widetilde{\epsilon}$ for different $M_{W'}$.
For this purpose, the event selection strategy can be different for different $M_{W'}$ and therefore, it can be further optimized according to $M_{W'}$.
For $M_{W'}=1 \;{\rm TeV}$, which is close to the mass of the $W$ boson in the SM, we use the looser cuts $p^{\ell}_T \geq 300\;{\rm GeV}$ and $\slashed{E}_T \geq 250\;{\rm GeV}$.

\begin{figure}[H]
\centering{
	{\includegraphics[width=0.49\textwidth]{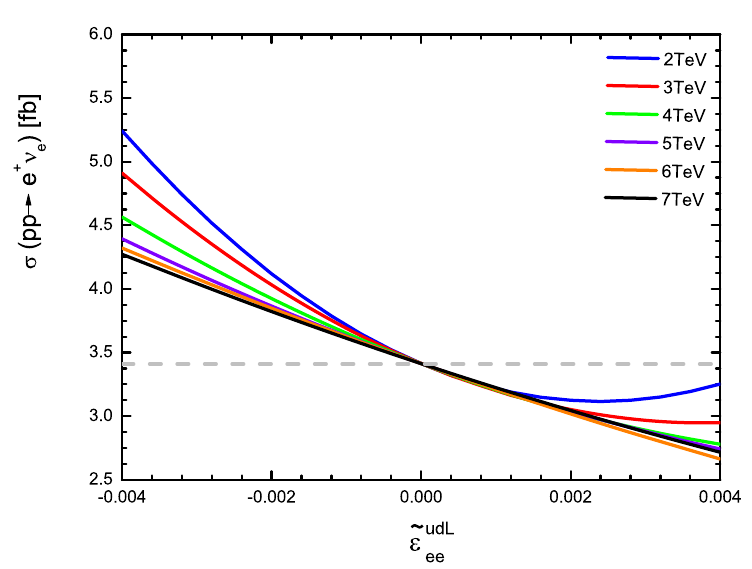}}
	{\includegraphics[width=0.49\textwidth]{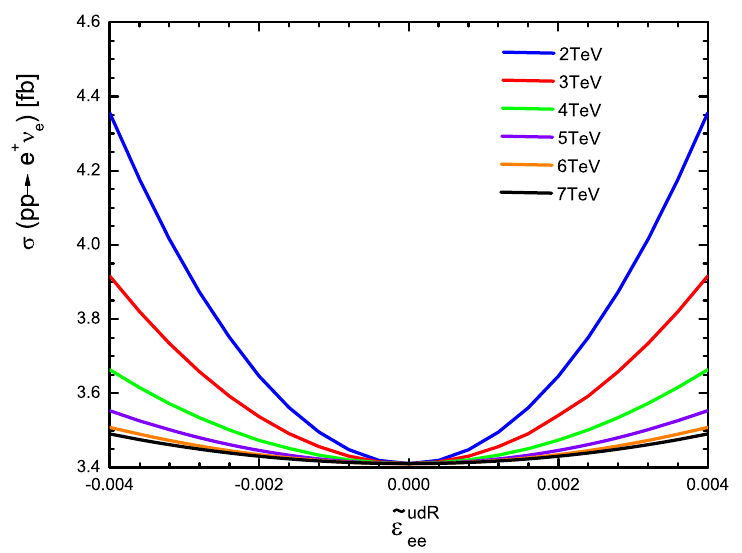}}
\caption{\label{fig:6} After the event selection strategy, the relationship between $\widetilde{\epsilon}^{u d L}_{e e}$ (left panel), $\widetilde{\epsilon}^{u d R}_{e e}$ (right panel), and the cross-section of the process $pp\to e^{+} \nu_{e}$ for different $M_{W'}$. The case of $M_W' = 1\;{\rm TeV}$ is not shown.}}
\end{figure}

Because the masses of fermions are negligible, the interference between $\widetilde{\epsilon} ^{qq' L}_{\ell\ell}$ and $\widetilde{\epsilon} ^{qq' R}_{\ell\ell}$ is neglected in this study for simplicity.
To focus on the sensitivities of the process on $\widetilde{\epsilon} $, we assume $\widetilde{\epsilon} $ is a real number in numerical estimations.
The cross-sections of the process $pp\to \ell^+\nu_{\ell}$ including the $W'$ contributions can be parameterized as
\begin{equation}
\sigma^L (\widetilde{\epsilon})= \sigma_{SM} + \sigma_{INT}(\widetilde{\epsilon}) + \sigma^L_{NSI}(\widetilde{\epsilon}),
\label{eq.antsaz7}
\end{equation}
\begin{equation}
\sigma^R (\widetilde{\epsilon})= \sigma_{SM} + \sigma^R_{NSI}(\widetilde{\epsilon}).
\label{eq.antsaz8}
\end{equation}
$\sigma_{INT}(\widetilde{\epsilon}) = \alpha_{int} \times \widetilde{\epsilon}$ originates from the interference between the SM and $W'$ contribution, and $\sigma^Y_{NSI}$ represents the contribution from only $W'$ exchanges with $\sigma^Y_{NSI}(\widetilde{\epsilon}) = \alpha^Y_{nsi} \times \widetilde{\epsilon}^{2}$.
After the event selection strategy, the dependencies of the factors $\alpha_{int}$ and $\alpha^Y_{nsi}$ on $M_{W'}$ can be fitted with the results of the MC simulation for different $M_{W'}$ at $\sqrt{s}$ = $13 \;{\rm TeV}$, which are listed in Table~\ref{Tab:experiments}.
For $M_W' = 2\sim 7 \;{\rm TeV}$, $\sigma_{SM}^{ee}$ = $0.0034 \;{\rm pb}$ and $\sigma_{SM}^{\mu\mu } $ = $0.0027 \;{\rm pb}$~($\sigma_{SM}^{ee}$ = $0.021 \;{\rm pb}$ and $\sigma_{SM}^{\mu\mu } $ = $0.027 \;{\rm pb}$ for $M_W' = 1\;{\rm TeV} $ owing to different event selection strategies).
Taking $pp\to e^+ \nu_{e}$ as an example, the cross-sections of the process $pp\to {e}^{+} \nu_{e}$ after cuts for different $M_{W'}$ are shown in Fig.~\ref{fig:6}.
The numerical results fit the bilinear functions in Eqs.~(\ref{eq.antsaz7}) and (\ref{eq.antsaz8}) very well.
Fig.~\ref{fig:6} shows that the interference plays an important role in $\sigma ^L(\widetilde{\epsilon})$.
Moreover, as shown Fig.~\ref{fig:6} and presented in the next section, the sensitivity of the process $pp\to \ell ^+\nu_{\ell}$ on $W'$ at the LHC has already reached the region where the interference effect should be considered.

\begin{table}[H]
	\centering
	\caption{\label{Tab:experiments} After the event selection strategy, $\alpha_{int}$ and $\alpha^Y_{nsi}$ are fitted for different $M_{W'}$ at $\sqrt{s}$ = $13 \;{\rm TeV}$.}
	\label{table1}
 \begin{tabular}{|p{1.2cm}<{\centering}|p{1.5cm}<{\centering}|p{1.5cm}<{\centering}|p{1.5cm}<{\centering}|p{1.5cm}<{\centering}|p{1.5cm}<{\centering}|p{1.5cm}<{\centering}|}
		\hline
		\multirow{1}{*}{$M_{W'}$} \multirow{1}{*}{(TeV)} & \multicolumn{3}{c|}{$pp \rightarrow {e}^{+} \nu_{e}$}&\multicolumn{3}{c|}{$pp \rightarrow {\mu}^{+} \nu_{\mu}$}\\
		\cline{2-7}
		
		& $\alpha_{int}$(pb) &$ \alpha_{nsi}^{L}$(pb) &$ \alpha_{nsi}^{R}$(pb) & $\alpha_{int}$(pb) & $\alpha_{nsi}^{L}$(pb) &$ \alpha_{nsi}^{R}$(pb) \\
		
		\hline
		1	& -3.29 	&6873.86  & 8878.72	&-3.18   &9127.30 & 8360.4 \\
		\hline
		2  &   -2.49   &5215.94  &5911.49  &-2.31   &5966.57   &5872.99 \\
		\hline
		3	&-2.45	&3225.12 &3164.77	&-2.43	&3618.67	&3518.60\\
		\hline
		4    &-2.23	&1621.66 &1584.89	&-2.14	&1754.81  &1668.15\\
		\hline
		5	&-2.07	&964.73 &893.93	&-2.07	&1030.03 &881.07\\
		\hline
		6	&-2.07	&649.12 &614.93	&-1.88	&669.24	&604.85\\
		\hline
		7	&-1.95	&508.87 &501.16	&-1.84	&511.24	&499.81\\
		\hline
	\end{tabular}
\end{table}

\begin{figure}[H]
\centering{
			\includegraphics [width=5cm] {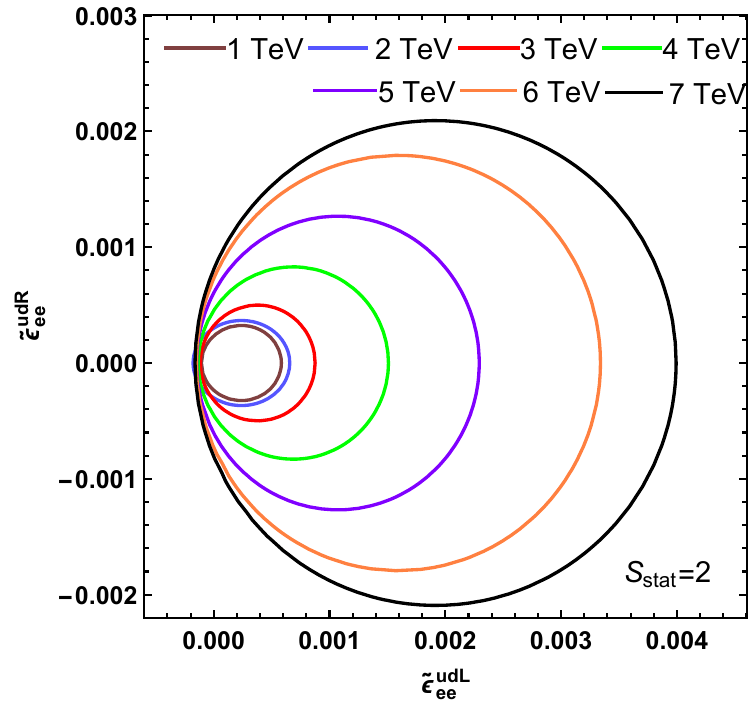}
			\includegraphics [width=5cm] {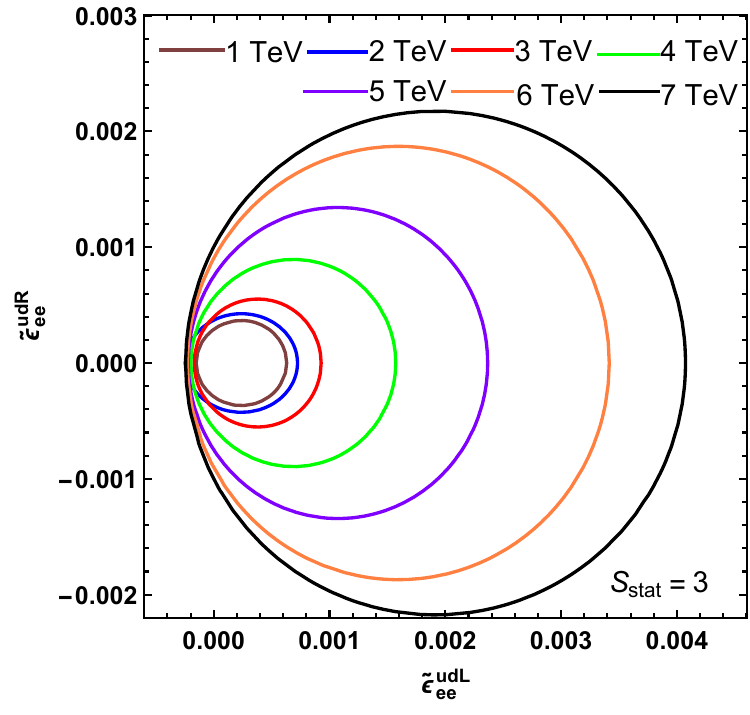}
			\includegraphics [width=5cm] {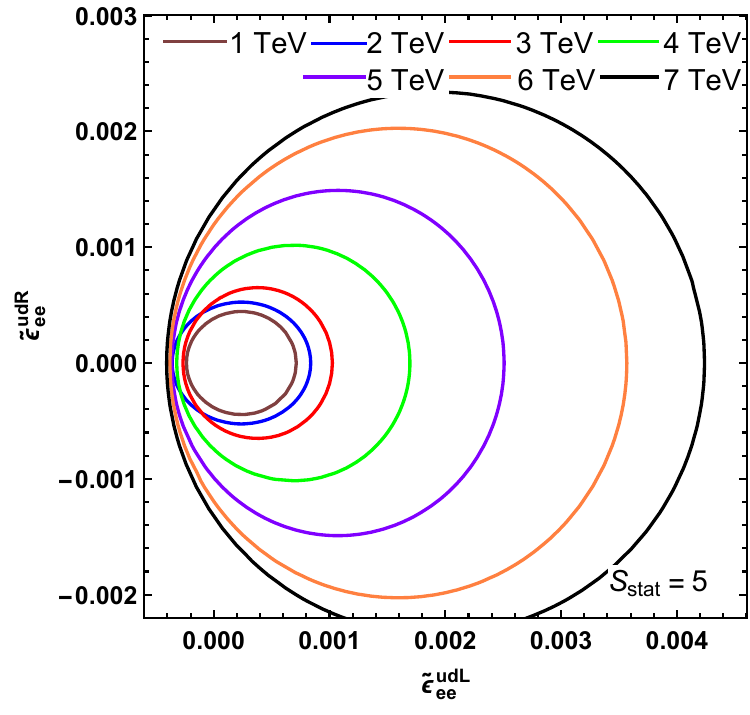}
	\caption{\label{fig:16}For $\mathcal{S}_{stat}$ equaling 2 (left panel), 3 (middle panel), and 5 (right panel), the relationship between the expected constraints on $\widetilde{\epsilon}^{udL}_{ee}$ and $\widetilde{\epsilon}^{udR}_{ee}$ at the $ 13 \;{\rm TeV}$ LHC with  $\mathcal{L}$ = $139\;{\rm fb}^{-1}$.   } }
\end{figure}

The expected constraints on $\widetilde{\epsilon}$ are estimated with the help of the statistical significance defined as
\begin{equation}
\mathcal{S}_{stat} = \sqrt{\mathcal{L}}\times \left( \left|\sigma\left(\widetilde{\epsilon}^{qq'Y}_{\ell\ell}\right)-\sigma _{SM}\right|/\sqrt{\sigma\left(\widetilde{\epsilon}^{qq'Y}_{\ell\ell}\right)} \right),
\label{eq.antsaz}
\end{equation}
where $\sigma _{SM}$ and $\sigma \left(\widetilde{\epsilon}^{qq'Y}_{\ell\ell}\right)$ are the SM cross-section and total cross-section including the $W'$ contributions after cuts are applied, respectively.
When $\widetilde{\epsilon}^{qq'L}_{\ell\ell}$ and $\widetilde{\epsilon}^{qq'R}_{\ell\ell}$ are both non zero, $\sigma (\widetilde{\epsilon}^L,\widetilde{\epsilon}^R) = \sigma_{SM} + \sigma_{INT}(\widetilde{\epsilon}^L) + \sigma^L_{NSI}(\widetilde{\epsilon}^L)+ \sigma^R_{NSI}(\widetilde{\epsilon}^R)$. For $\mathcal{S}_{stat}$ = 2, 3, and 5, the upper limits on the $\widetilde{\epsilon}^{udL}_{ee}$ and $\widetilde{\epsilon}^{udR}_{ee}$ plane are shown in Fig.~\ref{fig:16}.
For clarity, in Fig.~\ref{fig:16}, we use the signal significance of events that exceed the SM.
The constraints are approximate eccentric ellipses, indicating the importance of the interference term.

\begin{table}[H]
	\centering
	\caption{\label{Tab:conclusion} Expected constraints on  $\widetilde{\epsilon}^{udY}_{\ell\ell}$ detected at the $2\sigma$, $3\sigma$, and $5\sigma$ confidence levels (CLs).}
	\label{table1}
	 \begin{tabular}{|p{1.2cm}<{\centering}|p{3cm}<{\centering}|p{3cm}<{\centering}|p{3cm}<{\centering}|p{3cm}<{\centering}|}
		\hline		
		& $\widetilde{\epsilon}^{u d L}_{e e} (\times 10^{-4})$ &$ \widetilde{\epsilon}^{u d R}_{e e}(\times 10^{-4})$ &$ \widetilde{\epsilon}^{u d L}_{\mu \mu}(\times 10^{-4})$ & $\widetilde{\epsilon}^{u d R}_{\mu \mu}(\times 10^{-4})$  \\
		\hline
		$2\sigma$	& $ [-1.78,3.36 ]$	&$[-8.09,8.09]$ & $ [-1.77 , 3.35]$	&$[-7.31,7.31]$  \\
		\hline
		$3\sigma$  &  $ [-2.45,4.10]$   &$[-10.02,10.02]$ &$ [-2.43,4.09]$  &$[-9.55, 9.55]$ \\
		\hline
		$5\sigma$	&$ [-4.08,5.24]$	&$[-13.23,13.23]$ &$ [-3.92,5.20]$	&$[-12.61,12.61] $\\
		\hline
	\end{tabular}
\end{table}

\begin{figure}[H]
\centering{
	{\includegraphics[width=0.49\textwidth]{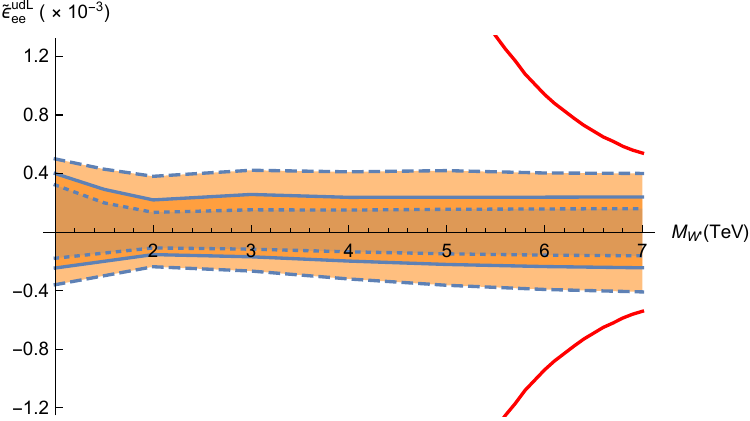}}
	{\includegraphics[width=0.49\textwidth]{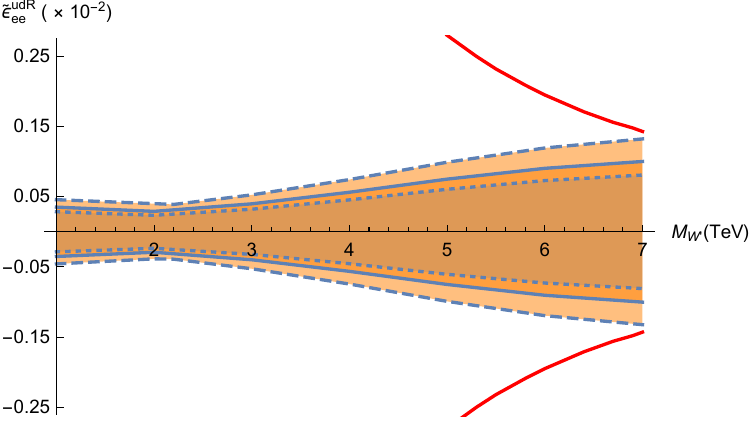}}
\caption{\label{fig:7}For $\mathcal{S}_{stat}$ equaling 2 (dotted line), 3 (solid line), and 5 (dashed line), the expected constraints on $\widetilde{\epsilon}^{udL}_{ee}$ (left panel) and $\widetilde{\epsilon}^{udR}_{ee}$ (right panel) as functions of $M_{W'}$  at the $ 13 \;{\rm TeV}$ LHC with $\mathcal{L}$ = $139\;{\rm fb}^{-1}$. The red solid lines originate from the tightest theoretical constraints.}}
\end{figure}

\begin{figure}[H]
\centering{
	{\includegraphics[width=0.49\textwidth]{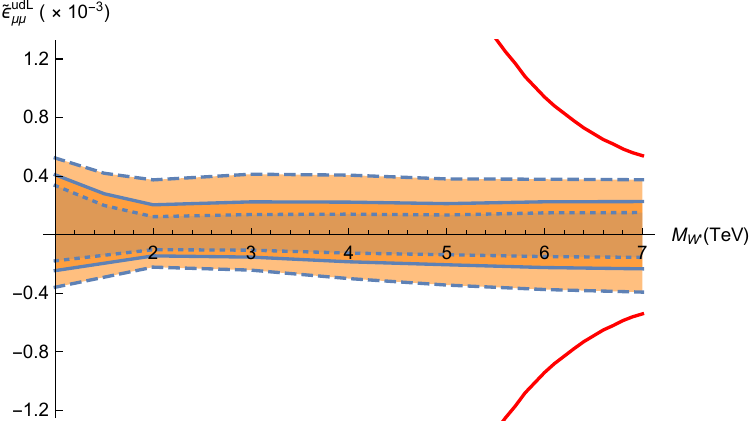}}
	{\includegraphics[width=0.49\textwidth]{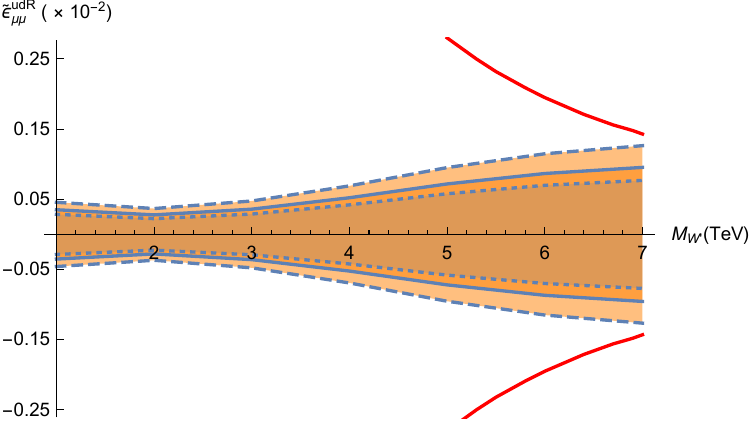}}
\caption{\label{fig:8}Same as Fig.~\ref{fig:7} but for $\widetilde{\epsilon}^{u d L}_{\mu \mu}$ (left panel) and $\widetilde{\epsilon}^{u d R}_{\mu \mu}$ (right panel).}}
\end{figure}

To study the sensitivities, it is assumed that only one of $\widetilde{\epsilon}^{qq'L}_{\ell\ell}$ and $\widetilde{\epsilon}^{qq'R}_{\ell\ell}$ is non zero. For $M_{W'}$ in the range of $ 1 \;{\rm TeV} \sim 7 \;{\rm TeV}$,
the expected constraints on $\widetilde{\epsilon}^{udY}_{\ell\ell}$ at the $2\sigma$, $3\sigma$, and $5\sigma$ confidence levels (CLs) are shown in Table~\ref{Tab:conclusion}.
In Figs.~\ref{fig:7} and~\ref{fig:8}, we show the expected constraints on $\widetilde{\epsilon}^{udL(R)}_{ee}$ and $\widetilde{\epsilon}^{u d L(R)}_{\mu \mu}$ for different $M_{W'}$ at the $13 \;{\rm TeV}$ LHC with $\mathcal{L}$ = $139\;{\rm fb}^{-1}$.
The tightest theoretical constraints discussed in Sec.~\ref{level2} are also presented in Figs.~\ref{fig:7} and~\ref{fig:8}.
The results of $\widetilde{\epsilon} ^{udY}_{\ell\ell}$ satisfy the unitarity bounds generated by  Eq.~(\ref{eq.boundY}) and satisfy the constraints given by Eq.~(\ref{eq.theoraticalConstraint}).
From Figs.~\ref{fig:7} and~\ref{fig:8}, we find that for different $M_{W'}$, the expected constraints are similar for $\widetilde{\epsilon} ^{ud L}_{\ell\ell}$.
This can be understood by looking at Table~\ref{Tab:experiments}; $\alpha _{int}$ are at the same order of magnitude.
Therefore, when the interference term is dominant, the expected constraints are insensitive to $M_{W'}$.
For the same reason, the positive and negative expected constraints are different for $\widetilde{\epsilon}^{udL}_{\ell\ell}$.
This is not the case for $\widetilde{\epsilon}^{udR}_{\ell\ell}$.
Because there is no interference, the expected constraints on $\widetilde{\epsilon} ^{ud R}_{\ell\ell}$ are generally looser than those on $\widetilde{\epsilon} ^{ud L}_{\ell\ell}$.
Except for the case of $M_{W'}=1\;{\rm TeV}$, the expected constraints are looser when $M_{W'}$ is larger because the cross-section decreases rapidly with increasing $M_{W'}$.
 For $M_{W'}=1\;{\rm TeV}$, the mass of the $W'$ boson is closer to that of the SM $W$ boson; therefore, the event selection strategy is less efficient. As a consequence,
even though the cross-section of the signal is larger, a stronger expected constraint cannot be reached compared with the case of $M_{W'}=2\;{\rm TeV}$.

\begin{figure}[H]
\centering{
		\includegraphics [width=12cm] {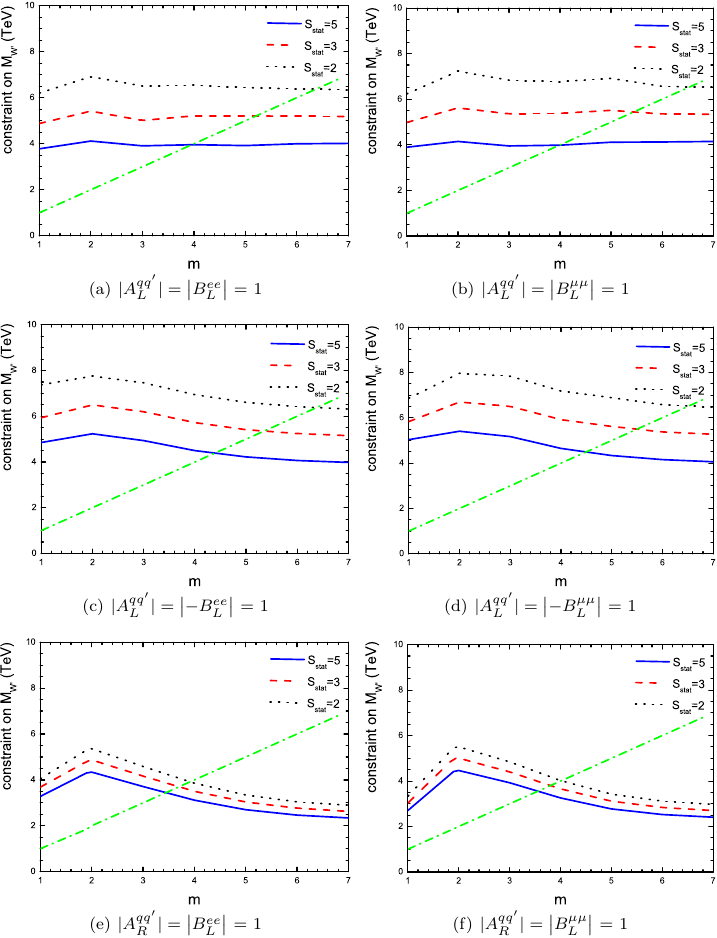}
	\caption{\label{fig:14}For $|{A}^{qq'}_{Y}| = \left|{B}_{L}^{\alpha\alpha}\right|$ = 1, the expected constraints on $M_{W'}$ obtained by events generated at $M_{W'} = m$. $M_{W'}$ on the left hand side of the dashed-dotted line is ruled out. The intersection of the dashed-dotted lines and the dotted lines gives the lower bound for $\mathcal{S}_{stat}$ = 2 as an example. }}
\end{figure}

It is interesting to discuss the expected constraints for special cases.
In the case of $|A^{qq'}_Y| = |B^{\alpha\alpha}_L|=1$, which corresponds to the SSM~\cite{Altarelli:1989ff}, the expected constraints on $\widetilde{\epsilon}$ can be translated to the lower bounds on $M_{W'}$.
The results are shown in Fig.~\ref{fig:14}.
However, because the MC simulation is carried out for fixed $M_{W'}$, if the lower bound on $M_{W'}$ is larger than the value of $M_{W'}$ used in the simulation, such value should be ruled out.
Consequently, an expected constraint can be set on $M_{W'}$.
The expected constraints of the process $pp\to e^+\nu _e$ on $M_{W'}$ for $\mathcal{S}_{stat}$ = 2, 3, and 5 are $M_{W'} > 6.3\;{\rm TeV}$, $M_{W'} > 5.2\;{\rm TeV}$, and $M_{W'} > 3.8\;{\rm TeV}$, respectively.
The expected constraints of the process $pp\to \mu^+\nu _{\mu}$ on $M_{W'}$ for $\mathcal{S}_{stat}$ = 2, 3, and 5 are $M_{W'} > 6.5\;{\rm TeV}$, $M_{W'} > 5.4\;{\rm TeV}$, and $M_{W'} > 3.9\;{\rm TeV}$, respectively.
The results of the LHC experiments are $M_{W'} > 6.0\;{\rm TeV}$ for the $e\nu _e$ channel and $M_{W'} > 5.0\;{\rm TeV}$ for the $\mu\nu _{\mu}$ channel at the $2\sigma$ CL (we choose the largest lower bounds from Refs.~\cite{ATLAS:2017jbq,CMS:2016ifc,CMS:2018hff,ATLAS:2019lsy}).
 By considering the interference effect, these results can be further improved.

\section{\label{level4} Sensitivities of the HL-LHC to \texorpdfstring{$\widetilde{\epsilon}^{udY}_{\ell\ell}$}{epsilon}}

\begin{figure}[H]
\centering{
	\includegraphics [width=11cm] {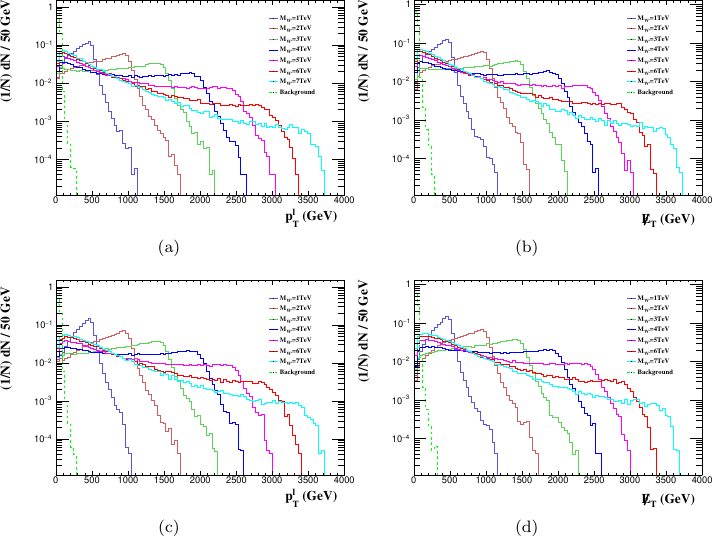}
\caption{\label{fig:9}Same as Fig.~\ref{fig:4} but for the HL-LHC at $\sqrt{s}$ = $14 \;{\rm TeV}$ with $\mathcal{L}$ = $300\;{\rm fb}^{-1}$.}}
\end{figure}

In this section, sensitivities to $\widetilde{\epsilon}^{udY}_{\ell\ell}$ in future runs of the LHC with higher luminosity, known as the HL-LHC, are investigated.
We assume that the HL-LHC runs at $\sqrt{s}$ = $14 \;{\rm TeV}$ with integrated luminosities of $300\;{\rm fb}^{-1}$, $1\;{\rm ab}^{-1}$, and $3\;{\rm ab}^{-1}$~\cite{Jezequel:2013mpa,Henderson:2021qrr}.
Taking the process $pp \rightarrow  e^{+} \nu_{e}$ as an example, the normalized distributions of $p_{T}^{\ell}$ and ${E\mkern-10.5mu/}_{T}$ for the signal and background are shown in Fig.~\ref{fig:9}.

\begin{figure}[H]
\centering{
	{\includegraphics[width=0.49\textwidth]{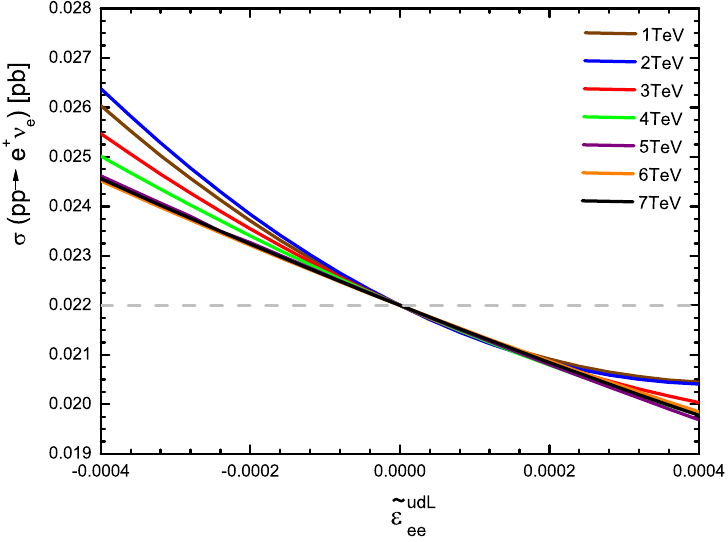}}
	{\includegraphics[width=0.49\textwidth]{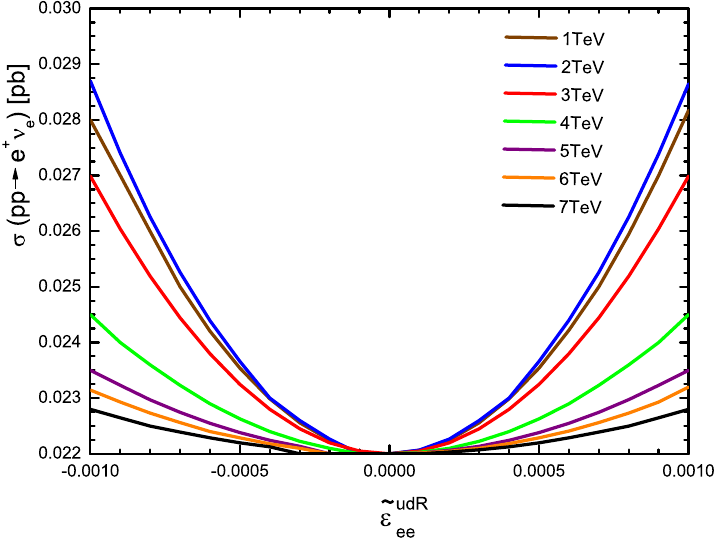}}
\caption{\label{fig:10} Same as Fig.~\ref{fig:6} but for the HL-LHC with $\sqrt{s}$ = $14 \;{\rm TeV}$.}}
\end{figure}

\begin{figure}[H]
\centering{
			\includegraphics [width=5cm] {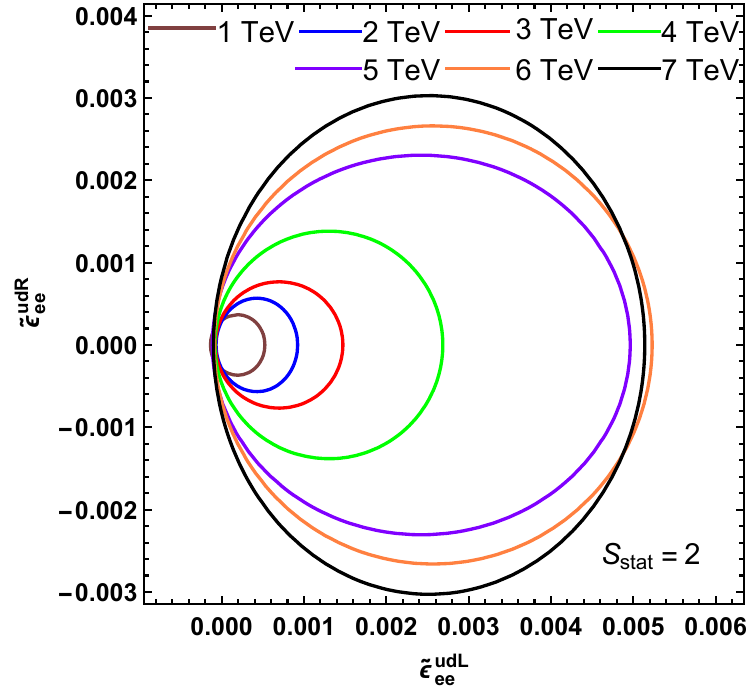}
			\includegraphics [width=5cm] {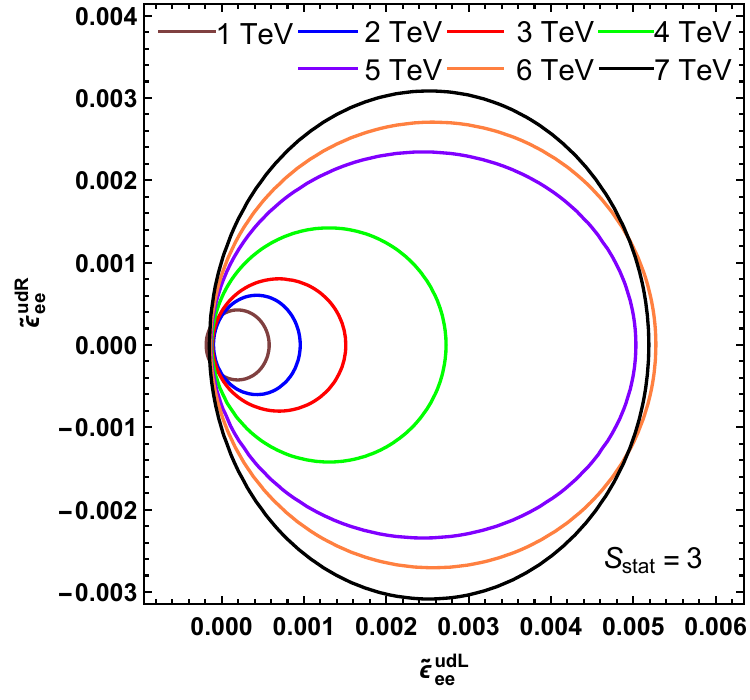}
			\includegraphics [width=5cm] {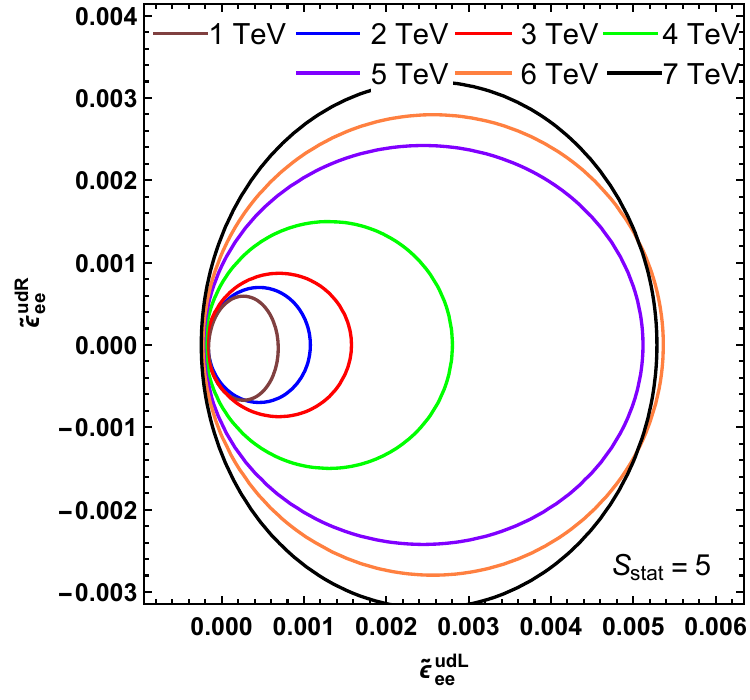}
	\caption{\label{fig:17}Same as Fig.~\ref{fig:16} but for the HL-LHC with $\sqrt{s}$ = $14 \;{\rm TeV}$. }}
\end{figure}

According to Fig.~\ref{fig:9}, for different masses ($1\;{\rm TeV} \sim 7\;{\rm TeV}$),
we choose the same cuts such that $p^{\ell}_T \geq 300\;{\rm GeV}$ and $\slashed{E}_T\geq 280\;{\rm GeV}$ for $pp\to {e}^{+} \nu_{e}$ and $p^{\ell}_T \geq 300\;{\rm GeV}$ and $\slashed{E}_T\geq 340\;{\rm GeV}$ for $pp\to {\mu}^{+} \nu_{\mu}$.
The cross-sections after cuts are also fitted using Eqs.~(\ref{eq.antsaz7}) and (\ref{eq.antsaz8}), $\sigma_{SM}^{ee}$ = $0.022\;{\rm pb}$, and $\sigma_{SM}^{\mu\mu }$ = $0.016 \;{\rm pb}$. The results are shown in Table~\ref{Tab:experiments2} and Fig.~\ref{fig:10}.
The upper limits on the $\widetilde{\epsilon}^{udL}_{ee}$ and $\widetilde{\epsilon}^{udR}_{ee}$ plane at the $14 \;{\rm TeV}$ LHC with $\mathcal{L}$ = $300\;{\rm fb}^{-1}$ are shown in Fig.~\ref{fig:17}.

\begin{table}[H]
	\centering
	\caption{\label{Tab:experiments2} Same as Table~\ref{Tab:experiments} but for the HL-LHC with $\sqrt{s}$ = $14 \;{\rm TeV}$.}
	 \begin{tabular}{|p{1.2cm}<{\centering}|p{1.5cm}<{\centering}|p{1.5cm}<{\centering}|p{1.5cm}<{\centering}|p{1.5cm}<{\centering}|p{1.5cm}<{\centering}|p{1.5cm}<{\centering}|}
		\hline
		\multirow{1}{*}{$M_{W'}$} \multirow{1}{*}{(TeV)} & \multicolumn{3}{c|}{$pp \rightarrow {e}^{+} \nu_{e}$}&\multicolumn{3}{c|}{$pp \rightarrow {\mu}^{+} \nu_{\mu}$}\\
		\cline{2-7}
		
		& $\alpha_{int}$(pb) &$ \alpha_{nsi}^{L}$(pb) &$ \alpha_{nsi}^{R}$(pb) & $\alpha_{int}$(pb) & $\alpha_{nsi}^{L}$(pb) &$ \alpha_{nsi}^{R}$(pb) \\
		
		\hline
		1	&-6.98 	&7782.23&6169.75	&-7.01	    &7171.98&6588.00    \\
		\hline
		2   &	-7.47 & 8729.47&6650.92  &-7.18&8159.10& 8507.71    \\
		\hline
		3	&-6.86	&4910.34&4990.20	&-6.76	&4752.89&5137.76	\\
		\hline
		4	&-6.54	&2510.25&2514.10	&-6.06	&2993.21&2888.60	\\
		\hline
		5	&-6.16	&1256.88&1522.74	&-5.68	&1227.32&1638.96	\\
		\hline
		6	&-5.89	&1146.44&1145.43	&-5.67	&1357.43&1120.93	\\
		\hline
		7	&-5.38	&1067.50&798.57	&-5.46	&680.62& 1054.37 	\\
		\hline
	\end{tabular}
\end{table}

\begin{figure}[H]
\centering{
	{\includegraphics[width=0.49\textwidth]{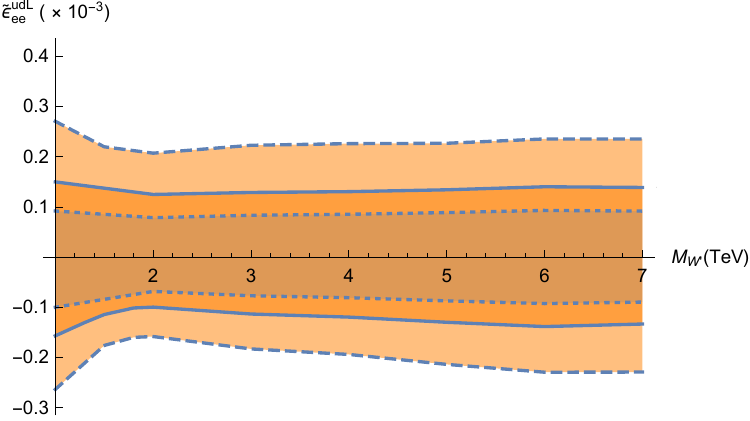}}
	{\includegraphics[width=0.49\textwidth]{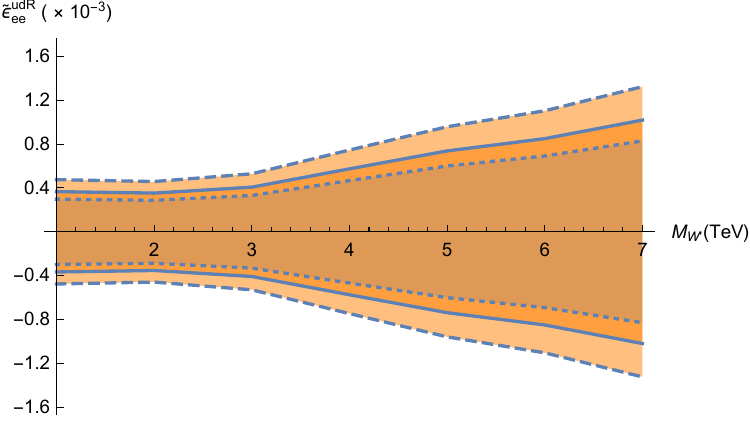}}
\caption{\label{fig:11}Same as Fig.~\ref{fig:7} but for $\widetilde{\epsilon}^{u d L}_{ee}$ (left panel) and $\widetilde{\epsilon}^{u d R}_{ee}$ (right panel) at the $14 \;{\rm TeV}$ HL-LHC with $\mathcal{L}$ = $300\;{\rm fb}^{-1}$.}}
\end{figure}

\begin{figure}[H]
\centering{
	{\includegraphics[width=0.49\textwidth]{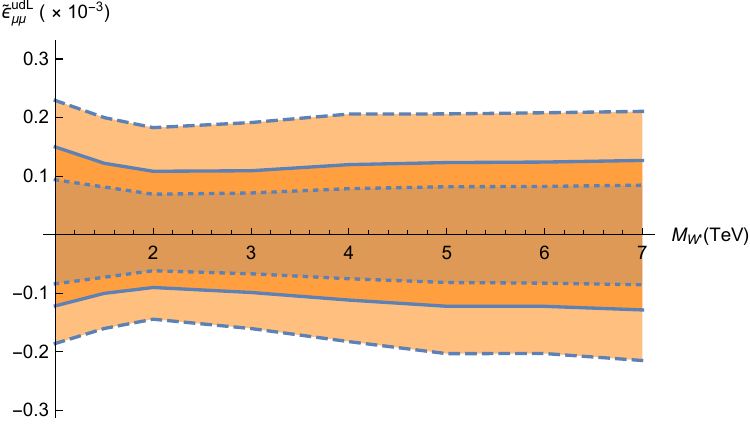}}
	{\includegraphics[width=0.49\textwidth]{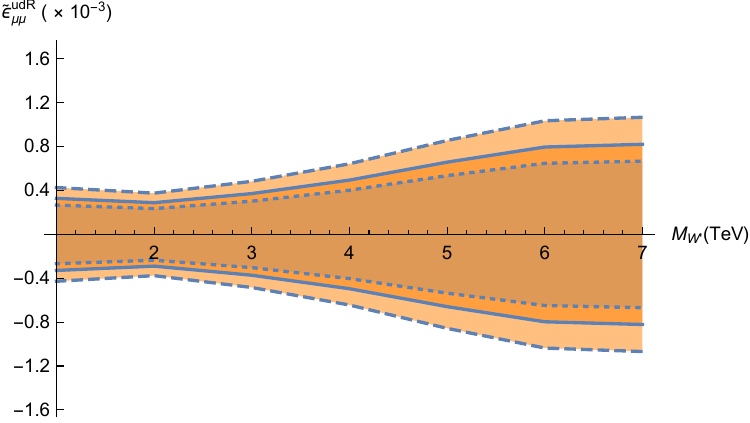}}
\caption{\label{fig:12}Same as Fig.~\ref{fig:11} but for $\widetilde{\epsilon}^{u d L}_{\mu \mu}$(left panel) and $\widetilde{\epsilon}^{u d R}_{\mu \mu}$(right panel).}}
\end{figure}

The sensitivities of the process $pp\to \ell^+ \nu_{\ell}$ at the $14 \;{\rm TeV}$ LHC with $\mathcal{L}$ = $300\;{\rm fb}^{-1}$ to $\widetilde{\epsilon}^{u d Y}_{\ell \ell}$ are estimated with the help of $\mathcal{S}_{stat}$. The numerical results are summarized in Figs.~\ref{fig:11} and~\ref{fig:12}.

\begin{figure}[H]
\centering{
	{\includegraphics[width=0.49\textwidth]{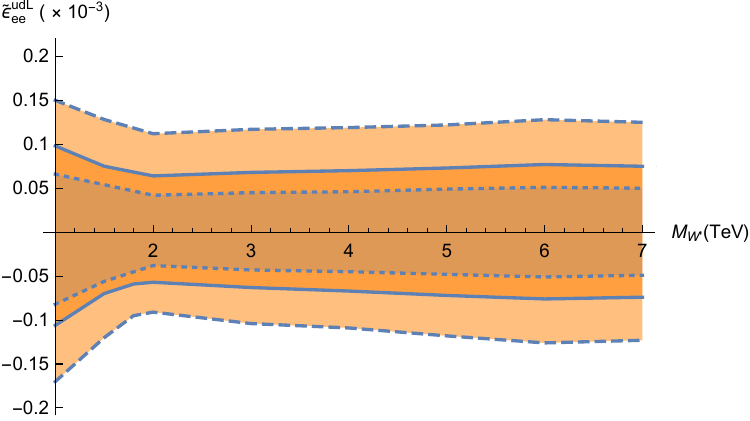}}
	{\includegraphics[width=0.49\textwidth]{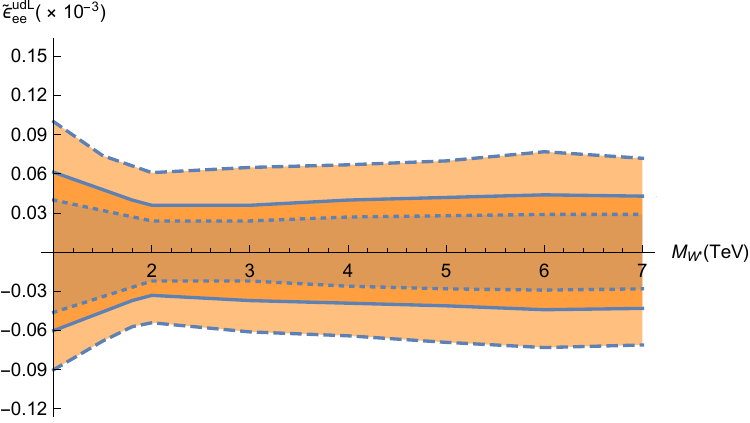}}
	\caption{\label{fig:15}Same as Fig.~\ref{fig:11} but for $\widetilde{\epsilon}^{u d L}_{ee}$ at the $14 \;{\rm TeV}$ LHC with $\mathcal{L}$ = $1\;{\rm ab}^{-1}$ (left panel) and $3\;{\rm ab}^{-1}$ (right panel).}}
\end{figure}

As shown in Figs.~\ref{fig:11} and~\ref{fig:12}, $\widetilde{\epsilon}^{u d Y}_{\ell\ell}$ satisfies the unitarity bounds generated by Eq.~(\ref{eq.boundY}) and is also within the constraints given by Eq.~(\ref{eq.theoraticalConstraint}).
Comparing these figures with Figs.~\ref{fig:7} and~\ref{fig:8}, the HL-LHC is more sensitive to the $W'$ boson than the LHC.
The results for higher luminosities, $\mathcal{L}$ = $1\;{\rm ab}^{-1}$ and $3\;{\rm ab}^{-1}$, are also investigated and shown in Fig.~\ref{fig:15}.
Compared with those at the $13\;{\rm TeV}$  LHC with $\mathcal{L}=139\;{\rm fb}^{-1}$, the expected constraints can be strengthened to approximately an order of magnitude for the $14\;{\rm TeV}$  LHC with  $\mathcal{L}=3\;{\rm ab}^{-1}$.

\section{\label{level5} Conclusions and discussions}

Many extensions of the SM predict the presence of charged heavy gauge bosons that might be found at the LHC, which are commonly referred to as the $W'$ boson and can induce the CC NSI.
In this paper, the contributions of $W'$ bosons to the process $pp\to \ell \nu$ are investigated.
Before MC simulation, we first consider the theoretical constraints on the simplified $W'$ model and further on $\widetilde{\epsilon}^{qq'Y}_{\ell\ell}$ from two perspectives: the partial wave unitarity and $W'$ decays.
Our numerical calculation shows that interference effects play a vital role, and as a consequence, the expected constraints on $\widetilde{\epsilon}^L$ are not less mass-dependent than in previous studies. The event selection strategy is then studied.
With the help of $\mathcal{S}_{stat}$, the expected constraints on $\widetilde{\epsilon}$ are estimated.
For the integrated luminosities considered in this paper, the expected constraints are insensitive to the nature of neutrinos (Dirac or Majorana)~\cite{Dib:2016wge}.

To date, there have been many studies on the constraints on NC NSI. For example, Ref.~\cite{Babu:2020nna} proposed that the LHC sensitivity to NC NSIs is $\widetilde{\epsilon}\leq 2\times10^{-3}$ for $M_{Z'}=2\;{\rm TeV}$, and the result of Ref.~\cite{Liu:2020emq} shows that in the simplified $Z'$ model, the upper limits on $\widetilde{\epsilon}$ are $0.042$ and $0.0028$ corresponding to $Z'$ with $M_{Z'}=0.2$ and $2\;{\rm TeV}$, respectively. However, up to our knowledge, studies on the new gauge boson $W'$ mainly focus on the constraints on its mass and couplings, and there are few studies on the CC NSI in the context of a simplified $W'$ model, especially at the current and future LHC. We focus on the expected constraints of collider experiments on the CC NSI induced by $W'$.
The expected constraints on the CC NSI parameters at the $2\sigma$ CL are $ -1.78\times 10^{-4} \leq \widetilde{\epsilon}^{u d L}_{e e} \leq 3.36 \times 10^{-4}$ and $ -1.77\times 10^{-4} \leq \widetilde{\epsilon}^{u d L}_{\mu \mu} \leq 3.35 \times 10^{-4}$ at the $13 \;{\rm TeV}$ LHC with $\mathcal{L}$ = $139\;{\rm fb}^{-1}$.
Ref.~\cite{Yue:2020wkj} carefully analyzed the contributions of $W'$ to everal low-energy observables, such as the leptonic decays of charged pion mesons, semileptonic $\tau$ decay, lepton flavor universality in pion mesons, CKM unitarity, and superallowed $\beta$ decays.
Among them, the tightest constraints are $\epsilon^{udV}_{ee}= 7.10\times10^{-4}$ from superallowed $\beta$ decays and $\epsilon^{udA}_{ee}= -1.85\times10^{-2}$ from the decay $\pi^{+} \rightarrow e^{+}\nu_{e}$, which result in $|\epsilon^{udL}_{ee}|< 8.895\times 10^{-3}$ and $|\epsilon^{udR}_{ee}|< 9.61\times10^{-3}$.
Thus, the expected constraints on the CC NSI parameters at the LHC are approximately an order of magnitude tighter than those from low-energy observables.
ATLAS and CMS presented the analysis of data with $L=36\;{\rm fb}^{-1}$ at $\sqrt{s}=13 \;{\rm TeV}$, setting the constraints as $M_{W'}>5.2\;{\rm TeV}$ in the electron channel~\cite{ATLAS:2017jbq} and $M_{W'}>4.9\;{\rm TeV}$ in the muon channel~\cite{CMS:2018hff}, at the $2\sigma$ CL using the SSM, which is looser than the constraints $6.3\;{\rm TeV}$ and $6.5\;{\rm TeV}$ in our paper.
Moreover, the expected constraints at the $14\;{\rm TeV}$ LHC with $\mathcal{L}=3\;{\rm ab}^{-1}$ can further narrow down the expected constraints to one order of magnitude from those at the $13\;{\rm TeV}$ LHC.
We propose that the interference effects are non-negligible and should be considered in future studies.

\section*{ACKNOWLEDGMENT}
This work was partially supported by the National Natural Science Foundation of China under Grant Nos. 11875157 and 12147214, the Natural Science Foundation of the Liaoning Scientific Committee No. LJKZ0978 and the Outstanding Research Cultivation Program of Liaoning Normal University (No. 21GDL004).

%

\appendix
\section{}
\setcounter{equation}{0}
\renewcommand{\theequation}{A.\arabic{equation}}
With the help of Eq.~(\ref{eq.sumrule}), the amplitudes corresponding to the Feynman diagrams in Fig.~\ref{fig:20} can be written as
\begin{align}
&\mathcal{M}\left(u_{-\frac{1}{2}}\bar{d}_{\frac{1}{2}}\to W'^{+}_{0} Z_{0}\right) =  g^{2} \left| V_{ud} \right|^{2} A^{ud}_{L} c_{W}\frac{\left(2M_{W'}^{2} + M_{Z}^{2}\right)}{2M_{W'}M_{Z}}e^{-i \phi} \sin\theta + \mathcal{O}(s^{-1}),\notag\\
&\mathcal{M}\left(d_{-\frac{1}{2}}\bar{u}_{\frac{1}{2}}\to W'^{-}_{0} Z_{0}\right) =  g^{2} \left|V_{ud} \right|^{2} A^{ud \ast}_{L} c_{W}\frac{\left(2 M_{W'}^{2} + M_{Z}^{2}\right)}{2 M_{W'}M_{Z}}e^{-i \phi}\sin\theta + \mathcal{O}(s^{-1}),\notag\\
&\mathcal{M}\left(u_{-\frac{1}{2}}\bar{u}_{\frac{1}{2}}\to W'^{+}_{0} W'^{-}_{0}\right) = -\left[ g^{2} \left| V_{ud} \right|^{2} \left| A^{ud}_{L} \right|^{2} - \left(\frac{1}{2}-\frac{2}{3}s_{W}^{2}\right)\frac{g^{2} \left| V_{ud} \right|^{2} M_{Z}^{2}}{2M_{W'}^{2}}\right.\notag \\
&- \left.\frac{g^{2} |V_{ud}|^{2} \left( |A^{ud}_{L}|^{2}-1\right)M_{Z'}^{2}}{4 M_{W'}^{2}} \right] e^{-i \phi} \sin\theta + \mathcal{O}(s^{-1}),\notag\\
&\mathcal{M}\left(d_{-\frac{1}{2}}\bar{d}_{\frac{1}{2}}\to W'^{+}_{0} W'^{-}_{0}\right) = \left[ g^{2} \left| V_{ud} \right|^{2} \left| A^{ud}_{L} \right|^{2} +\left(\frac{1}{3}s_{W}^{2}-\frac{1}{2}\right)\frac{g^{2} \left| V_{ud} \right|^{2}M_{Z}^{2}}{2M_{W'}^{2}}\right.\notag\\
&-\left.\frac{g^{2} \left| V_{ud} \right|^{2} (\left| A^{ud}_{L}\right|^{2}-1)M_{Z'}^{2}}{4 M_{W'}^{2}}\right]e^{-i \phi}\sin\theta +\mathcal{O}(s^{-1}),\notag\\
&\mathcal{M}\left(u_{-\frac{1}{2}}\bar{u}_{\frac{1}{2}}\to W'^{+}_{0} W^{-}_{0}\right) = \mathcal{M}\left(-d_{\frac{1}{2}}\bar{d}_{\frac{1}{2}}\to W'^{+}_{0} W^{-}_{0}\right)\notag \\
&= \pm g^{2} \left|V_{ud}\right|^{2} A^{ud}_{L}\left[  \frac{(M_{W'}^{2}+M_{W}^{2})}{2 M_{W'}M_{W}}-\frac{ M_{Z'}^{2}}{4 M_{W'}M_{W}}\right]e^{-i \phi}\sin\theta + \mathcal{O}(s^{-1}),\notag\\
&\mathcal{M}\left(u_{-\frac{1}{2}}\bar{u}_{\frac{1}{2}}\to W'^{-}_{0} W^{+}_{0}\right) = \mathcal{M}\left(-d_{\frac{1}{2}}\bar{d}_{\frac{1}{2}}\to W'^{-}_{0} W^{+}_{0}\right) \notag\\
&= \pm g^{2} \left|V_{ud}\right|^{2} A^{ud\ast}_{L}\left[ \frac{(M_{W'}^{2}+M_{W}^{2})}{2 M_{W'}M_{W}}+\frac{ M_{Z'}^{2}}{4 M_{W'}M_{W}}\right]e^{-i \phi}\sin\theta+ \mathcal{O}(s^{-1}),\notag\\
&\mathcal{M}\left(u_{\frac{1}{2}}\bar{d}_{-\frac{1}{2}}\to W'^{+}_{0} Z_{0}\right) = -g^{2} \left|V_{ud}\right|^{2} A^{ud}_{R}  s^{2}_{W}\frac{\left(2 M_{W'}^{2}+M_{Z}^{2}\right)}{2 c_{W}M_{W'}M_{Z}}e^{i \phi}\sin\theta + \mathcal{O}(s^{-1}),\notag\\
&\mathcal{M}\left(d_{\frac{1}{2}}\bar{u}_{-\frac{1}{2}}\to W'^{-}_{0} Z_{0}\right) = -g^{2} \left|V_{ud}\right|^{2} A^{ud\ast}_{R}  s^{2}_{W}\frac{\left(2 M_{W'}^{2}+M_{Z}^{2}\right)}{2 c_{W}M_{W'}M_{Z}}e^{i \phi}\sin\theta + \mathcal{O}(s^{-1}),\notag\\
&\mathcal{M}\left(u_{\frac{1}{2}}\bar{u}_{-\frac{1}{2}}\to W'^{+}_{0} W'^{-}_{0}\right) = -\left[g^{2} \left|V_{ud}\right|^{2} \left|A^{ud}_{R}\right|^{2} -\frac {g^{2}\left|V_{ud}\right|^{2}s_{W}^{4}M_{Z}^{2}}{3c^{2}_{W}M_{W'}^{2}}\right.\notag\\
&-\left.\left(\frac{\left|A^{ud}_{R}\right|^{2}}{2}-\frac{2s^{2}_{W}}{3c^{2}_{W}}\right)\frac{g^{2} \left|V_{ud}\right|^{2} M_{Z'}^{2}}{2M_{W'}^{2}}\right]e^{i \phi}\sin\theta + \mathcal{O}(s^{-1}),\notag\\
&\mathcal{M}\left(d_{\frac{1}{2}}\bar{d}_{-\frac{1}{2}}\to W'^{+}_{0} W'^{-}_{0}\right) = \left[g^{2} \left|V_{ud}\right|^{2} \left|A^{ud}_{R}\right|^{2}-\frac {g^{2}\left|V_{ud}\right|^{2}s_{W}^{4}M_{Z}^{2}}{6c^{2}_{W}M_{W'}^{2}}\right.\notag\\
&+\left.\left(\frac{s^{2}_{W}}{3c^{2}_{W}}-\frac{\left|A^{ud}_{R}\right|^{2}}{2}\right)\frac{g^{2} \left|V_{ud}\right|^{2}M_{Z'}^{2}}{2M_{W'}^{2}}\right]e^{i \phi}\sin\theta +\mathcal{O}(s^{-1}).\nonumber
\end{align}
\begin{align}
&\mathcal{M}\left(e^{-}_{-\frac{1}{2}}e^{+}_{\frac{1}{2}}\to W'^{+}_{0} W'^{-}_{0}\right) = \mathcal{M}\left(\mu^{+}_{-\frac{1}{2}}\mu^{-}_{\frac{1}{2}}\to W'^{+}_{0} W'^{-}_{0}\right)\notag \\
&=\left[g^{2}\left|B^{ee}_{L}\right|^{2}+\left(s_{W}^{2}-\frac{1}{2}\right)\frac {g^{2}M_{Z}^{2}}{2M_{W'}^{2}}-\frac {\left(|B^{ee}_{L}|^{2}-1\right)g^{2}M_{Z'}^{2}}{4M_{W'}^{2}}\right] e^{-i \phi}\sin\theta + \mathcal{O}(s^{-1}),\notag\\
&\mathcal{M}\left(e^{-}_{-\frac{1}{2}}e^{+}_{\frac{1}{2}}\to W'^{+}_{0} W^{-}_{0}\right) = \mathcal{M}\left(\mu^{+}_{-\frac{1}{2}}\mu^{-}_{\frac{1}{2}}\to W'^{+}_{0} W^{-}_{0}\right) \notag\\
&= g^{2}B^{ee\ast}_{L}\frac {2 \left(M_{W'}^{2}+M_{W}^{2}\right)-M_{Z'}^{2}}{4M_{W'}M_{W}}e^{-i \phi}\sin\theta + \mathcal{O}(s^{-1}),\notag\\
&\mathcal{M}\left(e^{-}_{-\frac{1}{2}}e^{+}_{\frac{1}{2}}\to W'^{-}_{0} W^{+}_{0}\right) = \mathcal{M}\left(\mu^{+}_{-\frac{1}{2}}\mu^{-}_{\frac{1}{2}}\to W'^{-}_{0} W^{+}_{0}\right) \notag\\
&= -g^{2}B^{ee}_{L}\frac {2 \left(M_{W'}^{2}+M_{W}^{2}\right)+M_{Z'}^{2}}{4M_{W'}M_{W}}e^{-i \phi}\sin\theta + \mathcal{O}(s^{-1}),\notag\\
&\mathcal{M}\left(e^{-}_{-\frac{1}{2}}\mu^{-}_{\frac{1}{2}}\to W'^{+}_{0} W'^{-}_{0}\right) = \mathcal{M}\left(\mu^{+}_{-\frac{1}{2}}e^{+}_{\frac{1}{2}}\to W'^{+}_{0} W'^{-}_{0}\right) = 2 g^{2}\left|B^{ee}_{L}\right|^{2}e^{-i \phi}\sin\theta + \mathcal{O}(s^{-1}),\notag\\
&\mathcal{M}\left(e^{-}_{-\frac{1}{2}}\mu^{-}_{\frac{1}{2}}\to W'^{+}_{0} W^{-}_{0}\right) = \mathcal{M}\left(\mu^{+}_{-\frac{1}{2}}e^{+}_{\frac{1}{2}}\to W'^{+}_{0} W^{-}_{0}\right)\notag \\
&=  g^{2}B^{ee\ast}_{R} \frac{M_{W'}^{2}+M_{W}^{2}}{2M_{W'}M_{W}}e^{-i \phi}\sin\theta + \mathcal{O}(s^{-1}),\notag\\
&\mathcal{M}\left(e^{-}_{-\frac{1}{2}}\mu^{-}_{\frac{1}{2}}\to W'^{-}_{0} W^{+}_{0}\right) = \mathcal{M}\left(\mu^{+}_{-\frac{1}{2}}e^{+}_{\frac{1}{2}}\to W'^{-}_{0} W^{+}_{0}\right) \notag\\
&=  -g^{2}B^{ee}_{L} \frac{M_{W'}^{2}+M_{W}^{2}}{2M_{W'}M_{W}}e^{-i \phi}\sin\theta + \mathcal{O}(s^{-1}),\notag\\
&\mathcal{M}\left(e^{-}_{-\frac{1}{2}}\bar{\nu_{e}}\to W'^{-}_{0} Z_{0} \right) =-\frac{ g^{2}B^{ee}_{L} c_{W}M_{Z}}{2 M_{W'}} e^{-i \phi}\sin\theta + \mathcal{O}(s^{-1}),\notag\\
&\mathcal{M}\left(\nu_{e} e^{+}_{\frac{1}{2}}\to W'^{+}_{0} Z_{0}\right)=\frac{ g^{2}B^{ee\ast}_{L} c_{W}\left(2 M^{2}_{W'}+M_{Z}^{2}\right)}{2 M_{W'}M_{Z}} e^{-i \phi}\sin\theta + \mathcal{O}(s^{-1}).
\label{eq.amplitudes}
\end{align}
$\mathcal{O}(s^{-1})$ indicates higher order terms. It is small and ignored in this study.

\begin{figure}[H]
	\centering
	\includegraphics [width=5cm] {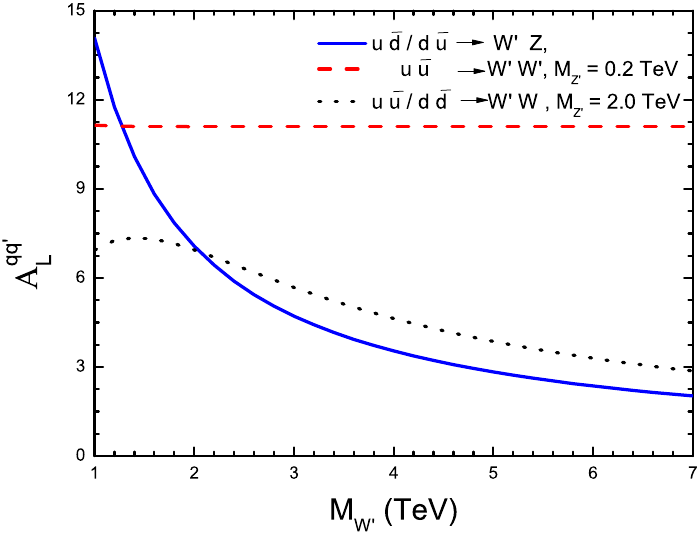}
	\includegraphics [width=5cm] {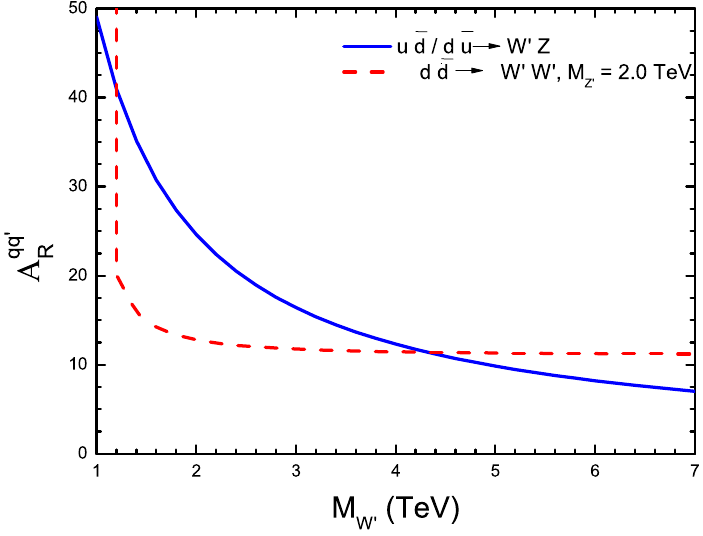}
	\includegraphics [width=5cm] {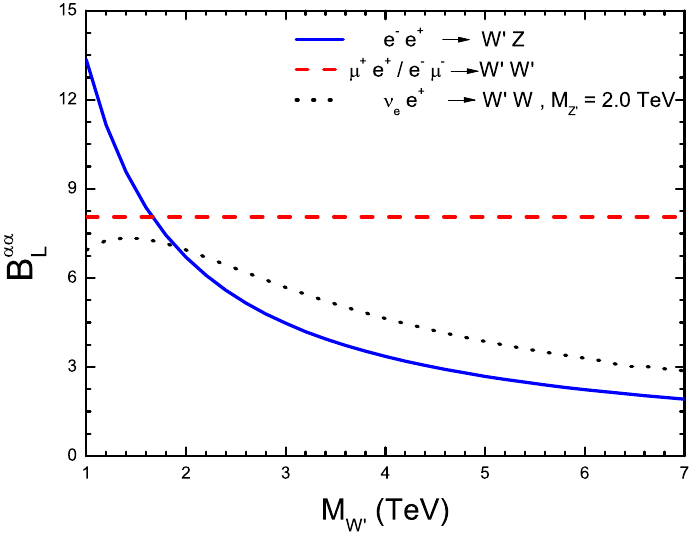}
\caption{\label{fig:01}Tightest constraints on the coupling parameters $A^{qq'}_{Y}$ and $ B^{\alpha\alpha}_{L}$ as functions of $M_{W'}$.  }
\label{figa}
\end{figure}

Using the relationship $|T_J|\leq 1$, we can obtain a constraint on ${A}^{qq'}_{Y}$ or ${B}^{\alpha\alpha}_{L}$ from each of the above processes. All of these constraints should be satisfied; therefore, we concentrate on the tightest ones, which depend on the free parameters $M_{W'}$ and $M_{Z'}$. In the case of $0.2\;{\rm TeV}<M_{Z'}<2\;{\rm TeV}$ ~\cite{Liu:2020emq} and $s\to \infty$, the tightest bounds are given by
\begin{align}
&{A^{ud}_{L}}\left(u_{-\frac{1}{2}}\bar{d}_{\frac{1}{2}}\to W'^{+}_{0} Z_{0}\right) ={A^{ud*}_{L}}\left(d_{-\frac{1}{2}}\bar{u}_{\frac{1}{2}}\to W'^{-}_{0} Z_{0}\right) <  \frac{24\sqrt{2}\pi M_{W'}M_{Z}}{g^{2} \left|V_{ud}\right|^{2} c_{W} \left(2 M_{W'}^{2}+M_{Z}^{2}\right)},\notag\\
&{A^{ud}_{L}}\left(u_{-\frac{1}{2}}\bar{u}_{\frac{1}{2}}\to W'^{+}_{0}W'^{-}_{0}\right) \notag\\&< \sqrt{\frac{48\sqrt{2} \pi M^{2}_{W'}-g^{2}\left|V_{ud}\right|^{2}M^{2}_{Z}+g^{2}\left|V_{ud}\right|^{2}M^{2}_{Z'}}{g^{2}\left|V_{ud}\right|^{2}\left|M^{2}_{Z'}-4M^{2}_{W'}\right|}+ \frac{4s^{2}_{W}M^{2}_{Z}}{3\left|M^{2}_{Z'}-4M^{2}_{W'}\right|}},\notag\\
&{A^{ud*}_{L}}\left(u_{-\frac{1}{2}}\bar{u}_{\frac{1}{2}}\to W'^{-}_{0}W^{+}_{0}\right)={A^{ud*}_{L}}\left(d_{-\frac{1}{2}}\bar{d}_{\frac{1}{2}}\to W'^{-}_{0}W^{+}_{0}\right) < \frac{48\pi M_{W'}M_{W}}{g^{2} \left|V_{ud}\right|^{2}\left(2 M_{W'}^{2}+2 M_{W}^{2}+ M_{Z'}^{2}\right)},\notag\\
&{A^{ud}_{R}}\left(u_{\frac{1}{2}}\bar{d}_{-\frac{1}{2}}\to W'^{+}_{0} Z_{0}\right) = {A^{ud*}_{R}}\left(d_{\frac{1}{2}}\bar{u}_{-\frac{1}{2}}\to W'^{-}_{0} Z_{0}\right) < \frac{24\sqrt{2}\pi c_{W}M_{W'}M_{Z}}{g^{2}\left|V_{ud}\right|^{2} s^{2}_{W} \left(2 M_{W'}^{2}+ M_{Z}^{2}\right)},\notag\\
&{A^{ud}_{R}}\left(d_{\frac{1}{2}}\bar{d}_{-\frac{1}{2}}\to W'^{+}_{0}W'^{-}_{0}\right) < \sqrt{\frac{48\sqrt{2}\pi M^{2}_{W'}}{g^{2}\left|V_{ud}\right|^{2}\left(4M^{2}_{W'}-M^{2}_{Z'}\right)} + \frac{2s^{2}_{W}\left(s^{2}_{W}M^{2}_{Z}-M^{2}_{Z'}\right)}{3c^{2}_{W}\left(4M^{2}_{W'}-M^{2}_{Z'}\right)}},\notag\\
&{B^{ee}_{L}}\left(e^{-}_{-\frac{1}{2}}e^{+}_{\frac{1}{2}}\to W'^{-}_{0} W^{+}_{0}\right) < \frac{48\sqrt{2}\pi M_{W'}M_{W}}{2g^{2} \left( M_{W'}^{2}+ M_{W}^{2}\right) +g^{2} M_{Z'}^{2}},\notag\\
& {B^{ee}_{L}}\left(e^{-}_{-\frac{1}{2}}\mu^{-}_{\frac{1}{2}}\to W'^{+}_{0} W'^{-}_{0}\right)= {B^{ee}_{L}}\left(\mu^{+}_{-\frac{1}{2}}e^{+}_{\frac{1}{2}}\to W'^{+}_{0} W'^{-}_{0}\right)  < \sqrt{\frac{6\sqrt{2}\pi}{g^{2}}},\notag\\
&{B^{ee*}_{L}}\left(\nu_{e} e^{+}_{\frac{1}{2}}\to W'^{+}_{0} Z_{0}\right)< \frac{24\sqrt{2}\pi M_{W'}M_{Z}}{g^{2}  c_{W}\left(2 M_{W'}^{2}+ M_{Z}^{2}\right)}.
\label{eq.coupling}
\end{align}
The relationship between $M_{W'}$ and the coupling parameters according to Eq.~(\ref{eq.coupling}) are shown in Fig.~\ref{figa}.
For a in wide range of $M_{W'}$, the tightest bound on ${A}^{qq'}_{L}$ originates from the process $u\bar{d}~(d\bar u)\to W'Z$. Similarly, the tightest bound on ${B}^{\alpha\alpha}_{L}$  mainly arises from the process $e^-e^+\to W'Z$. For simplicity, we only use the above bounds.
For ${A}^{qq'}_{R}$, the two tightest bounds~(depending on $M_{W'}$) are both considered.

\end{document}